\documentclass[usegraphicx,usenatbib,useapjfonts,apj]{emulateapj}

\slugcomment{}

\shorttitle{A catalogue of star formation and metallicity histories}
\shortauthors{Tojeiro et al.}

\begin{document}


\title{A public catalogue of stellar masses, star formation and metallicity histories and dust content from the Sloan 
Digital Sky Survey using VESPA}

\author{Rita Tojeiro\altaffilmark{1}}
\affil{Institute of Cosmology and Gravitation, Dennis Sciama Building, Burnaby Road, Portsmouth, PO1 3FX, UK}

\author{Stephen Wilkins}
\affil{Institute of Astronomy, University of Cambridge, Madingley Road, Cambridge, CB3 0HA, UK}

\author{Alan F. Heavens, Ben Panter}
\affil{Scottish Universities Physics Alliance (SUPA), Institute for Astronomy, University of Edinburgh, Blackford Hill, Edinburgh, EH9 3HJ, UK}
\and
\author{Raul Jimenez}
\affil{ICREA \& Institute of Space Sciences (CSIC-IEEC), Campus UAB, Bellaterra 08193, Spain}

\altaffiltext{1}{rita.tojeiro@port.ac.uk}

\begin{abstract}
We applied the VESPA algorithm to the Sloan Digital Sky Survey final data release of the Main Galaxies and Luminous Red Galaxies samples. The result is a catalogue of stellar masses, detailed star formation and metallicity histories and dust content of nearly 800,000 galaxies. We make the catalogue public via a T-SQL database, which is described in detail in this paper. We present the results using a range of stellar population and dust models, and will continue to update the catalogue as new and improved models are made public. The data and documentation are currently online, and can be found at http://www-wfau.roe.ac.uk/vespa/. We also present a brief exploration of the catalogue, and show that the quantities derived are robust: luminous red galaxies can be described by one to three populations, whereas a main galaxy sample galaxy needs on average two to five; red galaxies are older and less dusty; the dust values we recover are well correlated with measured Balmer decrements and star formation rates are also in agreement with previous measurements.

\end{abstract}



\section{Introduction}

The stellar mass of a galaxy has been shown to correlate with properties such as luminosity, morphology, star formation rate, mass density and stellar age, to name only a few (e.g. \citealt{ BrinchmannEllis00, BellEtAl03, HeavensEtAl04, BorchEtAl06, ShethEtAl06,BellEtAl07, ZhengEtAl07, PanterEtAl07}). Knowing how these relations evolve with redshift has been the goal of many observational studies, in an attempt to understand the main physical processes that drive star formation in galaxies. They can also provide strong constraints for models - these are normally ``tuned" for the local Universe, and seeing how well they predict the evolution of these quantities with redshift is a very powerful test. \\

Even though the stellar content of a galaxy is only the small tip of the
iceberg, it remains a very important component of the
Universe. Firstly because we can see it, and secondly because it holds
an imprint of that galaxy's star formation history, which combined with
other galaxies' provides information of when, how and where luminous
mass formed in the Universe.   \\

Galaxies' integrated colors alone can provide insight about their
evolution. The known bimodality of blue and red galaxies in a variety
of observables seems to tell us that these two populations are
intrinsically different. Whereas this is useful in its own right,
there is a considerable amount of more information to extract from
galactic light. In \cite{TojeiroEtAl07} we addressed the problem of
extracting information from a galaxy's integrated spectrum in a
reliable way, and presented VESPA as a tool to do so. In the current paper, we present the catalogue resulting from applying VESPA to the Sloan Digital Sky Survey's \citep{YorkEtAl00MNRAS, StraussEtAl02} final data release, which is now public and ready accessible.\\

First and foremost, this sort of analysis requires the means of physically interpreting
galactic light. A galaxy's spectrum can be modelled as a superposition
of stellar populations of different ages and metallicities, if we know
the expected flux of each stellar population. This is given by stellar
population models.  \\

Single stellar population models (SSPs) have three main ingredients. First we need a
description of the evolution of a star of given mass and metallicity in
terms of observable parameters, such as effective temperature and
luminosity (e.g. \citealt{AlongiEtAl93, BressanEtAl93, FagottoEtAl94a, GirardiEtAl96, MarigoEtAl08}. This can be calculated (or at least approximated)
analytically, to produce the so-called isochrones: evolutionary lines
for stars of constant metallicty in a color-magnitude
diagram. Secondly we need to assume an initial mass function (IMF),
which gives the number of stars per unit stellar mass, formed from a
single cloud of gas (e.g. \citealt{Salpeter55, Chabrier03, Kroupa07}). Different mass stars evolve with different
time-scales, and we can use the IMF to populate different evolutionary
stages of the color-magnitude diagram with the correct proportion of
stars of any given mass. Finally we need spectral libraries, which for
a combination of parameters such as luminosity or color index, assign
a spectrum to a star. Spectral libraries can either be drawn from our
local neighborhood, by taking high quality spectra of nearby stars (\citealt{LeBorgneEtAl03}),
or they can be theoretically motivated (e.g. \citealt{CoelhoEtAl05}).  \\

Stellar population models are limited in two main ways. Certain
advanced stages of stellar evolution, such as the supergiant phase, or
the asymptotic giant-branch phase, are hard to model and not always implemented. This
leads to uncertainties in the construction of the SSP models, which
are in this case worsened by the fact that these are bright stars
which contribute significantly to the overall luminous output. If
using empirical spectral libraries, stellar population models are also
limited by any bias of the stars in the solar neighborhood. For example, the
Milky Way is deficient in $\alpha$-elements (O, Ne, Mg, Si, S, Ca,
Ti), which are indicators of fast star formation. Nearby stars are
biased towards low [$\alpha$/Fe], which in turn bias the sample of high quality stellar
spectra available for collection. In this case
theoretical models might help, by explicitly calculating spectra for a
variety of [$\alpha$/Fe] models \citep{CervantesEtAl07, CoelhoEtAl07, LeeEtAl09}. For now, it should be understood that the metallicity implicit by the SSP models is [Fe/H], which is not degenerate with $\alpha$-element abundances.\\

The point to take home is that an analysis such as VESPA (and any others of the same type) is intrinsically model dependent, be it on the SSP modeling, IMF or dust modeling. The catalogue presented in this paper includes analyses done based on different combinations of models, and will continue to be updated as new models are released to the community, or as necessity demainds. It is not our immediate goal to distinguish which models better approximate the real Universe, but to provide the user with an opportunity to do so in their own studies. \\

\subsection{Extracting the information}

Extracting information from galactic spectra is a much more complex problem than that of extracting information from, for example, the cosmic microwave background's power spectrum. Firstly we must be clear about the parameters we want to extract from the data. We are faced with a non-trivial decision, since any parametrization we might choose will undoubtedly be an over-simplification of the problem - a galaxy is almost infinitely more complex than the early Universe. However, the quality of the data will often impose a limit on how many parameters we can safely recover from the data and one must be careful not to ask for more than what the data allows. The risk is getting back a solution which is largely dominated by noise, rather than real physics \citep{OcvirkEtAl06}. \\

From emission to absorption lines, continuum shape and spectral large-scale features, a galaxy's spectrum is packed with information about the physics of that galaxy. Stellar population and dust models provide us with a theoretical framework for their interpretation, and there are various ways in which one can do this. \\
 
Certain isolated spectral features are known to be well correlated with physical parameters, such as mass, star formation rate, mean age, or metallicity of a galaxy (e.g. \citealt{KauffmannEtAl03aMNRAS, TremontiEtAl04MNRAS, GallazziEtAl05, BarberEtAl06}). Absorption features are directly related to the chemical abundances of a stellar population, as they are created when the black-body emission from the centre of the star passes through its cooler outer regions. Certain absorption features, such as the Lick indices, have been well measured and calibrated so as to provide a standard set of tools which aid in assigning a physical meaning to a given absorption line (e.g. \citealt{Worthey94, ThomasEtAl03}). Emission lines are a sign of recent star formation: young, massive stars are the only ones with enough UV emission to ionize their surroundings. The recombination of the ionized gas creates signature emission lines, such as H$_\alpha$ and H$_\beta$, whose intensity (in the absence of dust) can tell us about the abundance of young stars in a galaxy. UV emission is, in itself, also a good probe for star formation for exactly the same reasons (e.g. \citealt{MadauEtAl96, Kennicutt98, HopkinsEtAl00, BrinchmannEtAl04, BundyEtAl06MNRAS, ErbEtAl06, AbrahamEtAl07MNRAS,NoeskeEtAl07MNRAS, SalimEtAl07, VermaEtAl07}).\\

VESPA focuses on using all of the available absorption features, as well as the shape of the continuum, in order to interpret a galaxy in terms of its star formation history. Emission lines are not included in the stellar population models (and are not present in every galaxy) and so we do not concentrate on these. Other methods have been developed to accomplish the same task: e.g., \cite{HeavensJimenezLahav00} (MOPED), \cite{CidFernandesEtAl04}  (STARLIGHT),  \cite{OcvirkEtAl06} (STECMAP), \cite{MacArthurEtAl09}, \cite{KolevaEtAl09} (ULySS). These and other methods
acknowledge the same limitation - noise in the data and in the models
introduces degeneracies into the
problem which can lead to unphysical results. Our approach with VESPA is to adapt the number of parameters needed to parametrize a galaxy to each  galaxy, taking into account the quality of the data. \cite{TojeiroEtAl07} showed how, by using the integrated spectrum of a galaxy, an appropriate parametrization can be found which recovers the maximum amount of information from a galaxy without running into the risk of over-parametrizing. \\

The result is a catalogue of robust and detailed star formation and metallicity histories, dust content and stellar mass for over 800,000 galaxies in SDSS's seventh data release, which we are now making public.\\

This paper is organized as follows: in Section \ref{sec:method} we briefly summarize our method (see \cite{TojeiroEtAl07} for full details) and the models we use; in Section \ref{sec:data} we summarize the data and pre-processing procedures; in Section \ref{sec:catalogue} we describe in detail each of the physical quantities output by VESPA; in Section \ref{sec:database} we lay out the technical details of the database and tables, and provide the user with some example queries for ease of access; in Section \ref{sec:results} we explore some properties of the catalogue and finally we present conclusions in Section \ref{sec:conclusion}. Where necessary, we assume a WMAP5 cosmology \citep{KomatsuEtAl08}.

\section{Method} \label{sec:method}

We use VESPA to analyze the seventh data release of the SDSS spectroscopic sample. The full details of our method can be found in \cite{TojeiroEtAl07}. For the sake of clarity and completeness however, we include here a summary that focuses on the issues that impact directly on how one might use the catalogue. \\

In short, VESPA solves the following problem:

\begin{equation}
\label{eq:vespa_problem}
F_{\lambda} = \int_0^t f_{dust}(\tau_\lambda, t) \psi(t) S_{\lambda}(t, Z)dt
\end{equation}

where $F_{\lambda}$ is the observed rest-frame of a galaxy, $\psi(t)$ is the star formation rate (solar masses formed per
unit of time) and $S_{\lambda}(t,Z)$ is the luminosity per unit
wavelength of a single stellar
population of age $t$ and metallicity $Z$, per unit mass. The
dependency of the metallicity on age is unconstrained, turning this into a
non-linear problem.\\

The problem consists in recovering the star formation history of the galaxy, and its dust content, by making assumptions about the form of $f_{dust}$ and $S_{\lambda}(t, Z)$. VESPA accomplishes this by writing the problem in a linear form, and minimizing 

\begin{equation}
\chi^2 = \frac{ \sum_{j} ( F_j - F^{recovered}_j)}{\sigma_j^2 }
\end{equation}

where $j$ represents a given bin in wavelength. Even though the problem has an analytical solution, a dataset perturbed by noise or which is otherwise deteriorated leads to instabilities in the matrix inversion and the recovered solutions can be entirely dominated by noise \citep{OcvirkEtAl06}. VESPA has a self-regularization mechanism which estimates how many independent parameters one should recover from a given a dataset that has been perturbed. The result is a parametrization which varies from galaxy to galaxy, depending on its signal-to-noise ratio (SNR) and wavelength coverage. \\

The overall goal of VESPA is to retrieve a solution which is robust, rather than very detailed in look-back time - we sacrifice precision for the sake of accuracy.\\

\subsection{VESPA's bins}
\begin{figure*}[htbp]
\begin{center}
\plotone{bins_table.epsf}
\caption[Schematic view of the grid of bins used by VESPA.]{Schematic view of the grid of bins used by VESPA. The top line of black numbers indicates the age of each boundary, in Gyrs. The red numbers in each of the bins is an unique bin identifier number, which can be used to quickly retrieve properties of a given bin.  }
\label{fig:bins_table}
\end{center}
\end{figure*}

\begin{figure*}[htbp]
\begin{center}
\plottwo{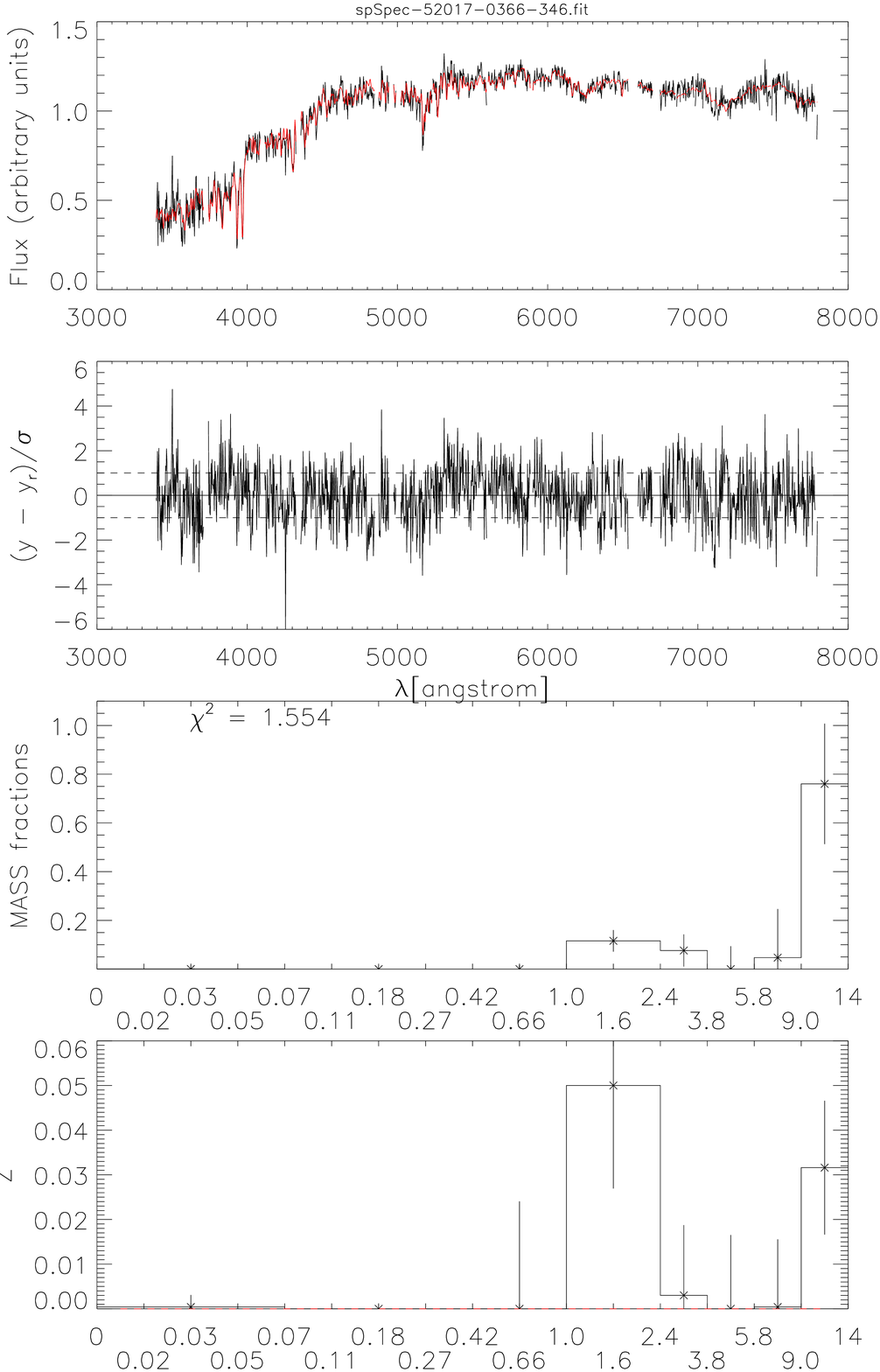}{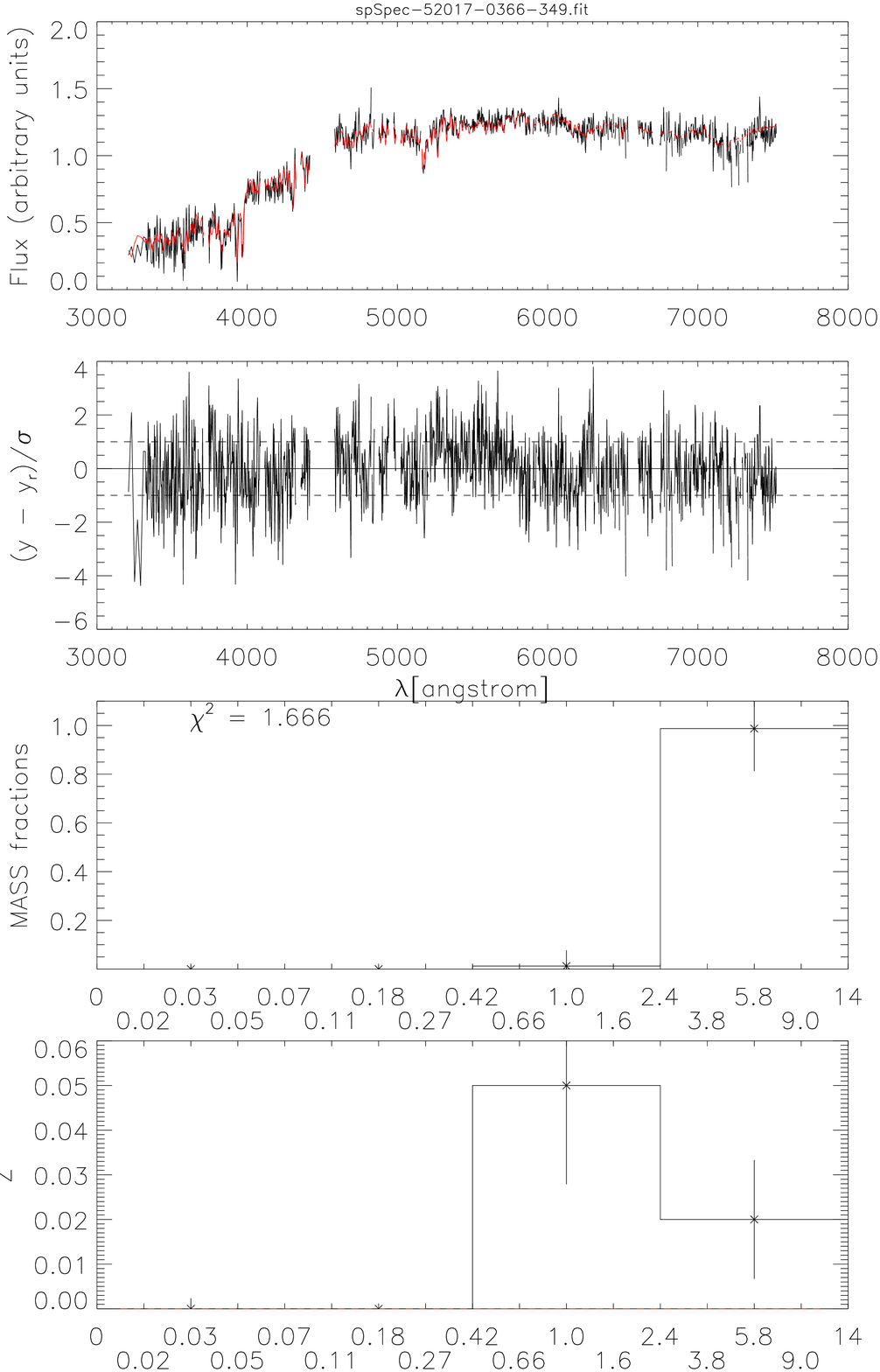}
\vspace{0.1in}
\caption{Two SDSS galaxies analysed with VESPA. In each case the top panels show the observed and fit spectrum (black and red, respectively; only the fitted regions are shown), the second panel the residuals, the third panel the recovered star formation mass fractions and in the bottom panel we show the recovered metallicity in each age bin. The example on the right shows a galaxy from which little information could safely be recovered which is translated into large age bins. The interpretation should be that the majority of this galaxy's mass was formed 11-14 Gyrs ago in the rest-frame, but we cannot tell more precisely when, within that interval, this happened. The example on the left shows a galaxy with a history which is better resolved.}
\label{fig:examples}
\end{center}
\end{figure*}

VESPA recovers stellar mass fractions for each age bin $\alpha$, which span a range of look-back time $\Delta t_\alpha$. The age of the Universe is split into different-sized bins, as detailed in Figure \ref{fig:bins_table}. In this figure we show a schematic view of the VESPA bins, together with the unique bin identifier number used in the published catalogue. The output for each galaxy generally consists of a combination of high-resolution bins populated with non-zero star formation, low resolution populated bins, and empty bins. Figure \ref{fig:examples} show the results for two galaxies with very different bin configurations. \\

Next we describe how we compute the SSP model flux of each of these bins, given a set of models $S(\lambda, t, Z)$.

\subsubsection{High-resolution age bins}
\label{sec:HRbins}
At our highest resolution (HR) we work with 16 age bins, equally spaced in a logarithmic
time scale between 0.002 Gyr and the age of the Universe. In each bin, we assume a
constant star formation rate

\begin{equation}
f^{HR}_\alpha(\lambda, Z) = \psi \int_{\Delta t_\alpha}
S(\lambda, t, Z) dt
\label{eq:HR_SFR}
\end{equation}
\par\noindent
with 
\begin{equation}
\psi = 1/\Delta t_\alpha.
\label{eq:HR_SFR}
\end{equation}

\subsubsection{Low-resolution age bins}
\label{sec:LR_bins}
We work on a grid of different resolution time
bins and we construct the low resolution bins (LR) using the high
resolution bins described in Section \ref{sec:HRbins}. We do not assume a constant star
formation rate in this case, as in wider bins the light from the
younger components would largely dominate over the contribution from
the older ones. Instead, we use a decaying star formation history,
such that the light contributions from all the components are
comparable. We start by writing
\begin{equation}
f^{LR}_\alpha(\lambda, Z) = \int_{\Delta t_\alpha} \psi(t) S(\lambda, t, Z) dt, 
\end{equation}
which we approximate to
\begin{equation}
f^{LR}_\beta(\lambda, Z) = \frac{\sum_{\alpha \in \beta} f^{HR}_\alpha(\lambda, Z) \psi_\alpha \Delta t_\alpha}{\sum_{\alpha \in \beta} \psi_\alpha \Delta t_\alpha}
\end{equation}
where low resolution bin $\beta$ incorporates the high resolution bins $\alpha \in
\beta$, and we set
\begin{equation}
\psi_\alpha \Delta t_\alpha = \frac{1}{\int_\lambda f^{HR}_\alpha(\lambda, Z)
  d\lambda}.
\label{eq:weights}
\end{equation}
\par\noindent

\subsection{Models}

VESPA can work with any set of SSP, IMF and dust models, and the solutions it recovers are inevitably model-dependent. \\

\subsubsection{SSP Modeling}
The modeling of SSPs is still very much an active and developing field. We chose to publish the catalogue using more than one set of SSP models to give the user the opportunity to check how their results fare with different models. At the time of this writing, we are publishing results obtained with the models of \cite{BruzualEtCharlot03} (BC03) and \cite{Maraston05} (M05). As different sets of models become available to the community, or as observational data demands, we intend to update the catalogue accordingly. \\

With the BC03 models we adopt a Chabrier initial mass
function \citep{Chabrier03} and Padova 1994 evolutionary tracks
\citep{AlongiEtAl93, BressanEtAl93, FagottoEtAl94a,
FagottoeEtAl94b, GirardiEtAl96}. We interpolate metallicities using tabulated values at: $Z=$ 0.0004, 0.004, 0.008, 0.02 and 0.05. We use the red horizontal branch models of \cite{Maraston05} with a Kroupa initial mass function \citep{Kroupa07}. Models are supplied at metallicities of $Z=$ 0.0004, 0.01, 0.02 and 0.04. Both models are normalized to $1M_\odot$ at $t=0$.\\

\subsubsection{Dust Modeling}\label{sec:dust_modeling}
\begin{figure}
\plotone{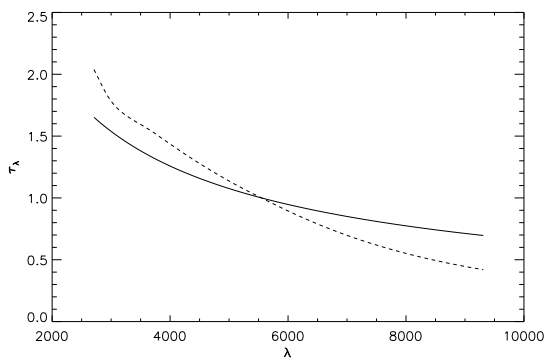}
\caption{Two commonly used dust extinction curves. The solid line shows
  a simple model that follows $\lambda^{-0.7}$ and is used throughout
  this paper. The dashed line shows the  the extinction
curve estimated directly from the Large Magellanic Cloud
  by \cite{GordonEtAl03}. Curves have been normalised to unity at
  $\lambda=5550\AA$. }
\label{fig:dust_curves}
\end{figure}

The simplest approach to dust modeling is to assume that stars of all ages are affected in the same way:

\begin{equation}
F_{\lambda} = f_{dust}(\tau_\lambda) \int_0^t \psi(t) S_{\lambda}(t, Z)dt.
\end{equation}

where $\tau_\lambda$ is the optical depth as a function of wavelength, and which we model as

\begin{equation}
\tau_\lambda =  \tau_V \left( \frac{\lambda}{5500 \AA} \right)^{-0.7}.
\end{equation}
A comparison of this curve with the estimated extinction curve of
the LMC by \cite{GordonEtAl03} can be seen in Figure
\ref{fig:dust_curves}. In this figure both curves are normalised
such that $\tau_{5550 \AA} = 1$. 

There is a variety of choices for the form of
$f_{dust}(\tau_\lambda)$. We use the mixed slab model of
\cite{CharlotFall00} for low optical depths ($\tau_V \le 1$), for which
\begin{equation}
f_{dust}(\tau_\lambda) = \frac{1}{2\tau_\lambda}[1 + (\tau_\lambda -
1)\exp(-\tau_\lambda) - \tau_\lambda^2E_1(\tau_\lambda)]
\end{equation}
where $E_1$ is the exponential integral and
$\tau_\lambda$ is the optical depth of the slab. This model is known
to be less accurate for high dust values, and for optical depths
greater than one we take a uniform screening model with
\begin{equation}
f_{dust}(\tau_\lambda) = \exp(-\tau_\lambda).
\end{equation}

We call this model our one-parameter dust model. We also apply a two-parameter dust model by following \cite{CharlotFall00}
and set

\begin{equation}
f_{dust}(\tau_\lambda, t) = \left\{
\begin{array}{l}
f_{dust}(\tau^{ISM}_\lambda) f_{dust}(\tau^{BC}_\lambda), t \leq t_{BC}\\
f_{dust}(\tau^{ISM}_\lambda), t > t_{BC}\\
\end{array}
\right.
\end{equation}

where $\tau^{ISM}$ refers to the inter-stellar medium and $\tau^{BC}$ to the birth cloud. This allows stellar populations yonger than $t_{BC}$ to have extra extinction in relation to the older populations. We only use the uniform screening model to model the dust in the birth
cloud and we use $\tau_\lambda =\tau_V(\lambda/5500\AA)^{-0.7}$ as our
extinction curve for both environments. We set $t_{BC}=0.03$ Gyr.\\
\par\noindent
As described, dust is a non-linear problem. In practice, we solve the linear
problem described by equation (\ref{eq:vespa_problem}) with a number of dust
extinctions applied to the models $S(t,Z)$ and choose the
values of $\tau_V^{ISM}$ and $\tau_V^{BC}$ which result in the best
fit to the data. We initially use a
binary chop search for $\tau_V^{ISM} \in [0,4]$ and keep $\tau_V^{BC}$
fixed and equal to zero, which results in trying out
typically around nine values of $\tau_V^{ISM}$. If this initial
solution reveals star formation at a time less than $t_{BC}$ we repeat
our search on a two-dimensional grid, and fit for $\tau_V^{ISM}$ and
$\tau_V^{BC}$ simultaneously. There is no penalty except in CPU time to apply the two-parameter search, but we find that this procedure is robust \citep{TojeiroEtAl07}.\\

\subsection{Mock galaxies and wavelength range} \label{sec:wavelength_range}

Any full spectral fitting code is dependent on the wavelength range which is fitted and in \cite{TojeiroEtAl07} we showed that an increased coverage allows for a more precise and more resolved solution. Here we take the opportunity to take this analysis further by explicitly examining how the wavelength shift due to redshift affects the recovered solutions. \\

We take a similar approach to \cite{TojeiroEtAl07} and simulate mock galaxies with two types of star formation histories: an exponentially decaying star formation rate ($SFR \propto \gamma t_\alpha$ and $\gamma = 0.3$ Gyr $^{-1}$), and a dual-burst history. We add Gaussian white noise for a SNR per $3\AA$ pixel of 50, and we take a fixed Earth-frame wavelength coverage of $[3800, 9200] \AA$, and excluding emission-line regions, as detailed in Section \ref{sec:handling_data}. In all cases the metallicity is assigned randomly for each age. \\

We assess the quality of the recovered solutions in terms of the star formation history, metallicity history, and total stellar mass. Here we depart from our previous methodology and calculate goodness-of-fit estimates, in solution space, taking into account the recovered errors (see Section \ref{sec:errors} for details on how these are calculated). To avoid making the database too large we only keep the diagonal components of the covariance matrices, so we approximate $\chi^2$ by:
\begin{equation}
\chi^2_{SFH} = \sum_\alpha \frac{(m_\alpha^{I} - m_\alpha^{R})^2}{C_{\alpha,\alpha}(m)}
\end{equation}

where $m_\alpha^I$ and $m_\alpha^R$ are the input and recovered mass in bin $\alpha$, and $C(x)$ is the covariance matrix for the mass (see Section  \ref{sec:errors}). We compute $\chi^2_{Z}$ in an identical fashion.  To assess how well we recover stellar mass we simply calculate $(M^I - M^R)/\sigma_M$, where $\sigma_M$ is given by the mass covariance matrix, and $M^I$ and $M^R$ are the input and recovered total stellar mass, respectively. \\

We expect $\chi^2_{SFH}$ and $\chi^2_Z$ to be distributed like a $\chi^2$ distribution with $n$ degrees of freedom. The complication, however, is that $n$ varies from galaxy to galaxy, as the number of bins changes. Given that we are dealing with uniform samples of galaxies, we do not expect this variation to be very large but it makes more sense to compute a reduced value of $\chi^2_{SFH}$ and $\chi^2_{Z}$ in order to make the distributions more uniform . We compare the obtained distributions with the expected $\chi^2$ distribution with $5$ degrees of freedom, as this is the average number of populations recovered.\\ 

\begin{figure*}[htbp]
\begin{center}
\plotone{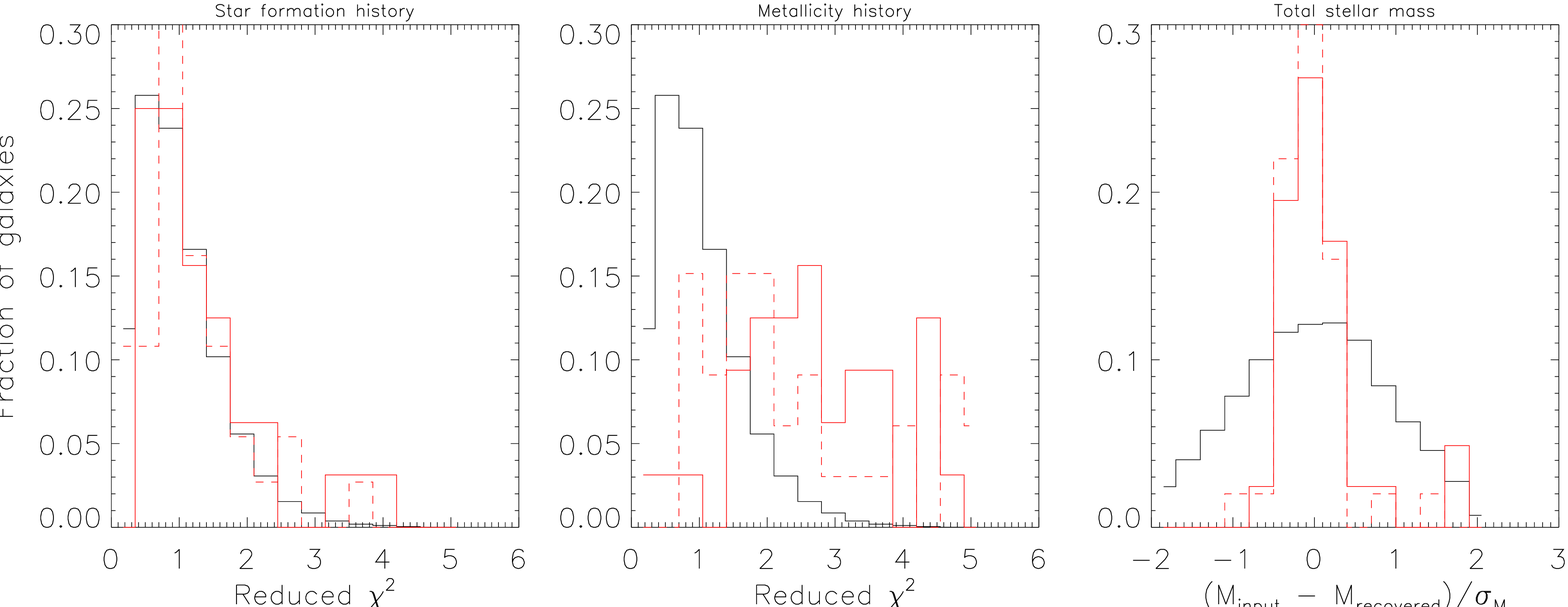}
\vspace{0.1in}
\caption{$\chi^2_{SFH}, \chi^2_{Z}$ and $(M^I - M^R)/\sigma_M$ for 50 mock galaxies with an exponentially decaying star formation history, at a redshift of 0.05 (red solid line) and 0.2 (red dashed line). For reference, we also show the expected distributions in the same binning scheme (black solid line). See main text for discussion.}
\label{fig:mocks_tau03_nerror50}
\end{center}
\end{figure*}

Figures \ref{fig:mocks_tau03_nerror50} and \ref{fig:mocks_dual_nerror50} show these distributions for a set of 50 mock galaxies at redshifts of 0.05 and 0.2, with an exponentially decaying and a dual-burst SFH respectively. \\

\begin{figure}[htbp]
\begin{center}
\plotone{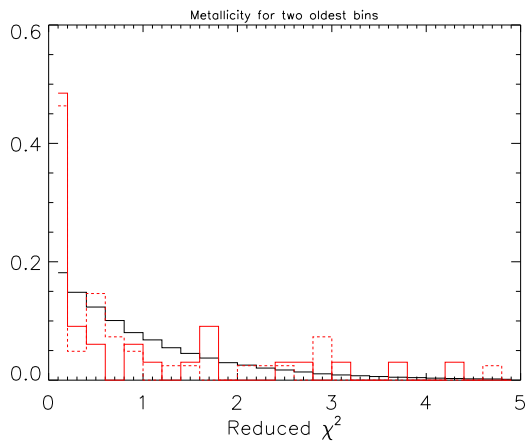}
\vspace{0.1in}
\caption{The distribution of $\chi^2_Z$ for the two oldest bins only, in mock galaxies with an exponentially decaying star formation history, at a redshift of 0.05 (red solid line) and 0.2 (red dashed line). For reference, we also show the expected distribution in the same binning scheme (black solid line). The contrast between this distribution and that shown in the middle panel of Figure \ref{fig:mocks_tau03_nerror50} shows how metallicities of populations which have a low contribution in mass/light is highly unconstrained. }
\label{fig:metallicity_old_tau}
\end{center}
\end{figure}

\begin{figure*}[htbp]
\begin{center}
\plotone{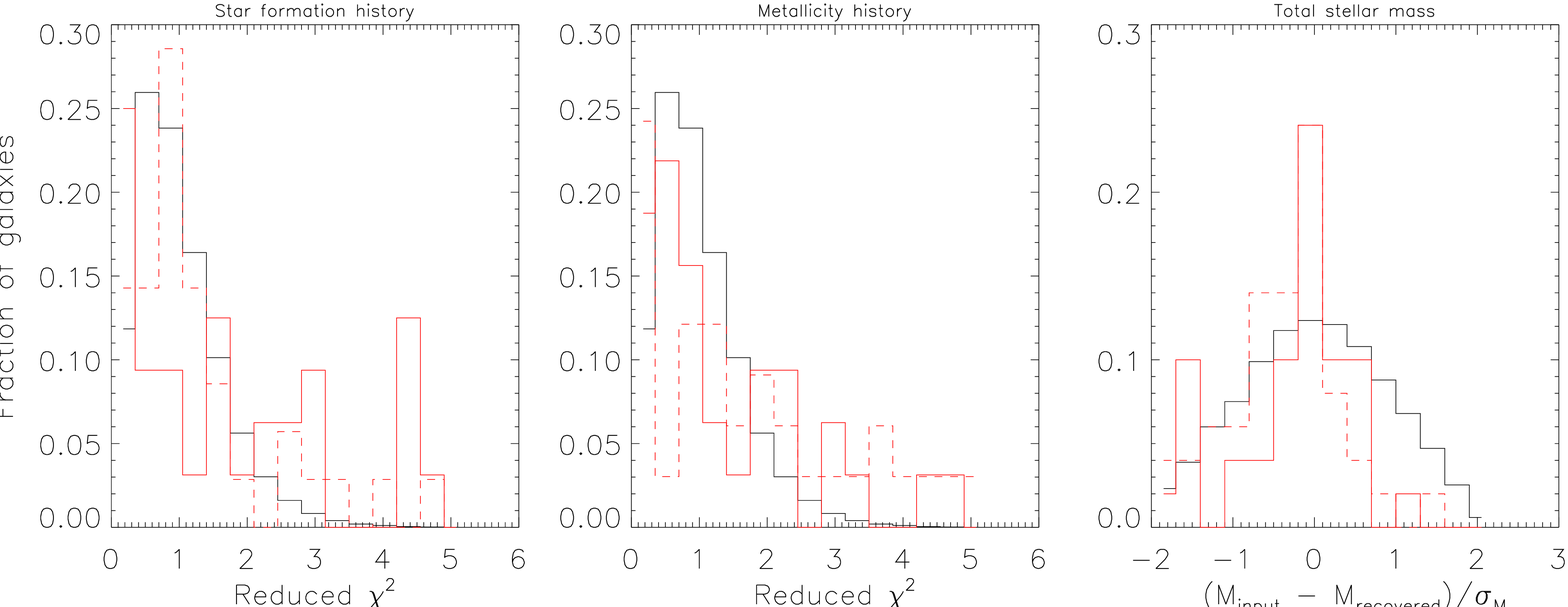}
\vspace{0.1in}
\caption{$\chi^2_{SFH}, \chi^2_{Z}$ and $(M^I - M^R)/\sigma_M$ for 50 mock galaxies with a dual-burst history, at a redshift of 0.05 (red solid line) and 0.2 (red dashed line). For reference, we also show the expected distributions in the same binning scheme (black solid line). See main text for discussion.}
\label{fig:mocks_dual_nerror50}
\end{center}
\end{figure*}

In the exponentially decaying case, we find that recovered star formation history is perfectly consistent with the expected distribution, meaning that the recovered solutions and their errors are well estimated. In the case of metallicity, however, the recovered distributions seem incompatible. This is dominated by the fact that in most cases, the stellar mass formed in each bin is very low (with the exception of the two oldest), which makes it very hard to accurately recover the metallicity. If we restrict our analysis to the two oldest bins, the resulting $\chi^2_Z$ is much more pleasing as we show in Figure \ref{fig:metallicity_old_tau} - here the expected distribution is computed for $n=2$. Figure \ref{fig:metallicity_old_tau} suggests that perhaps the error on those two bins are {\it over-estimated}, but the recovered solutions are excellent. The middle panel in Figure \ref{fig:mocks_tau03_nerror50} is a stark reminder that recovered metallicities for populations which contribute very little to the light of the galaxy are poorly constrained, as we pointed out in \cite{TojeiroEtAl07}. The total stellar mass is well recovered, although we see signs that our total error is an over-estimated by roughly a factor of two. The error is determined from the scatter of the recovered solutions of a set of mock galaxies, each generated with a SFH identical to that recovered, but with different noise.  The SFH is not the true SFH of the galaxy (since this is not available in the real world), and this leads to an error estimate which is somewhat overestimated. In practice, however, this is not significant in real galaxies, where other factors are more important; namely the modelling and spectro-photometric calibrations (see e.g. Figure 14 of \cite{TojeiroEtAl07}). \\

Finally, we note that there is no significant difference in the recovered solutions as a function of redshift.\\

\begin{figure}[htbp]
\begin{center}
\plotone{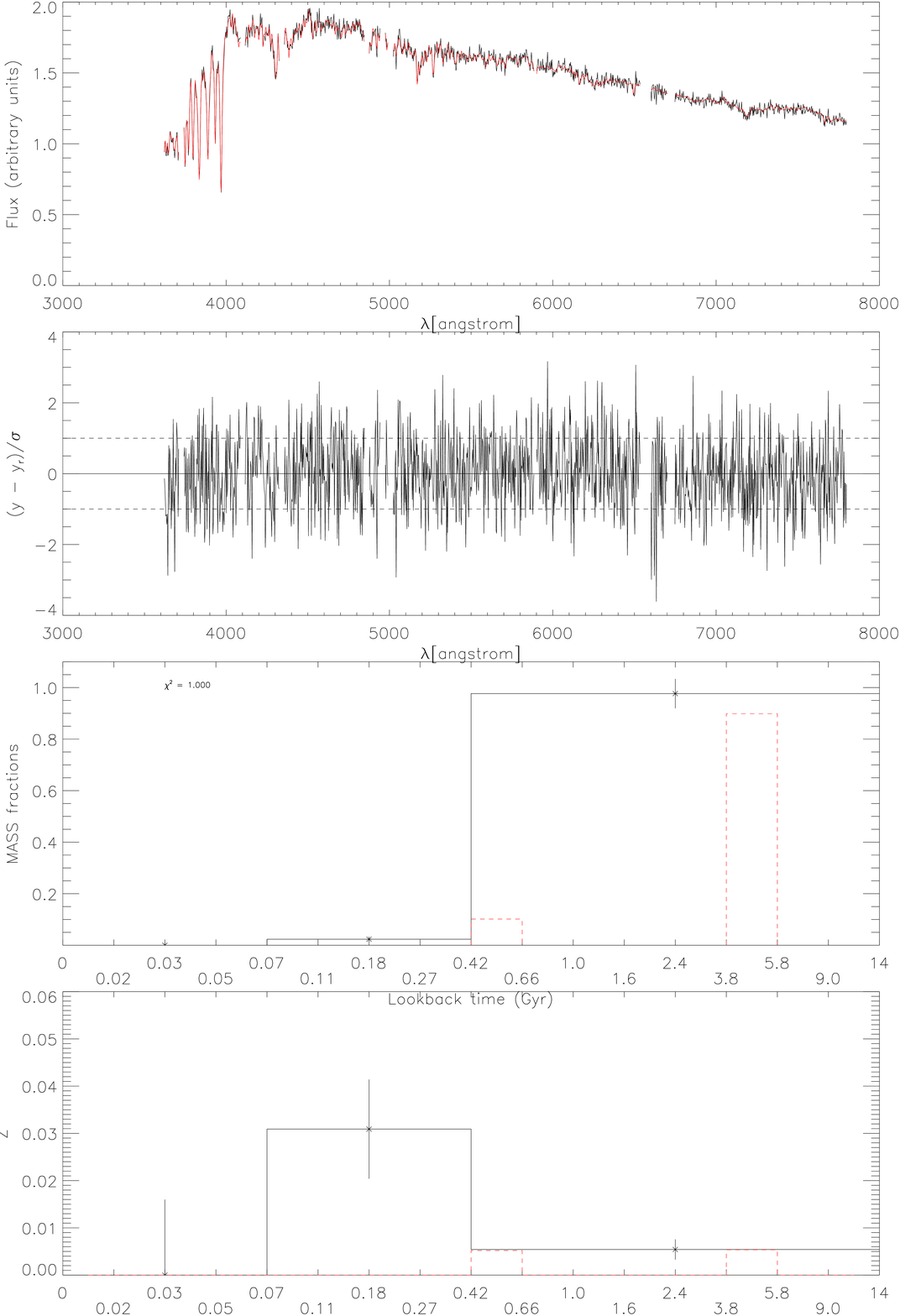}
\vspace{0.1in}
\caption{An example of a mock galaxy where the $\chi_{SFH}^2$ is unlikely poor, in spite of the good fit in data space. In this case, this is due to the wide bin and the assumed star formation rate - see text for details.}
\label{fig:mock_dual_example}
\end{center}
\end{figure}

In case of a dual-burst history, as shown in Figure \ref{fig:mocks_dual_nerror50}, we see that in general the recovered star formation history is not as well recovered, given the errors. There is a wider variety in the $n$ from galaxy to galaxy, so the differences for a low-value of reduced $\chi_{SFH}$ and $\chi_{Z}$ are not worrying, but the excess of galaxies with poor values of goodness-of-fit is. Some of these objects also have an unlikely poor goodness-of-fit in data space, which is always an indication that the solutions are poor, but the majority does not. The latter, however, have very wide bins. Take the example in Figure \ref{fig:mock_dual_example}. The recovered {\it mass fractions} (plotted) are accurate, but the recovered {\it absolute mass} is not. This is because there is an implied star formation rate within each bin that ultimately determines the mass-to-light ratio of that bin, which in turn gives the the absolute mass recovered. The wider the bin, the more important this assumption becomes. Therefore, one should be careful when interpreting histories which are very poorly resolved. The metallicities tend to be better recovered in this case because there are fewer age bins with very low star formation. The scatter in total stellar mass is larger, but still good.\\

Once again we fail to see any systematic offset due to the different redshifts of the mock galaxies. \\

\subsubsection{Emission-line regions}

A different question concerns the effect that removing emission-line regions has on the recovered solutions. As will be detailed in Section \ref{sec:handling_data}, we remove certain parts of the rest-frame wavelength range which correspond to common emission lines. We do this for all galaxies, irrespective of color, morphology or detection of emission lines by the SDSS pipeline, and have also removed them from our mock analysis in the previous section. However, some of the {\it absorption} features in these regions can be important for the recovery of the solutions. \\

\begin{figure*}[htbp]
\begin{center}
\plotone{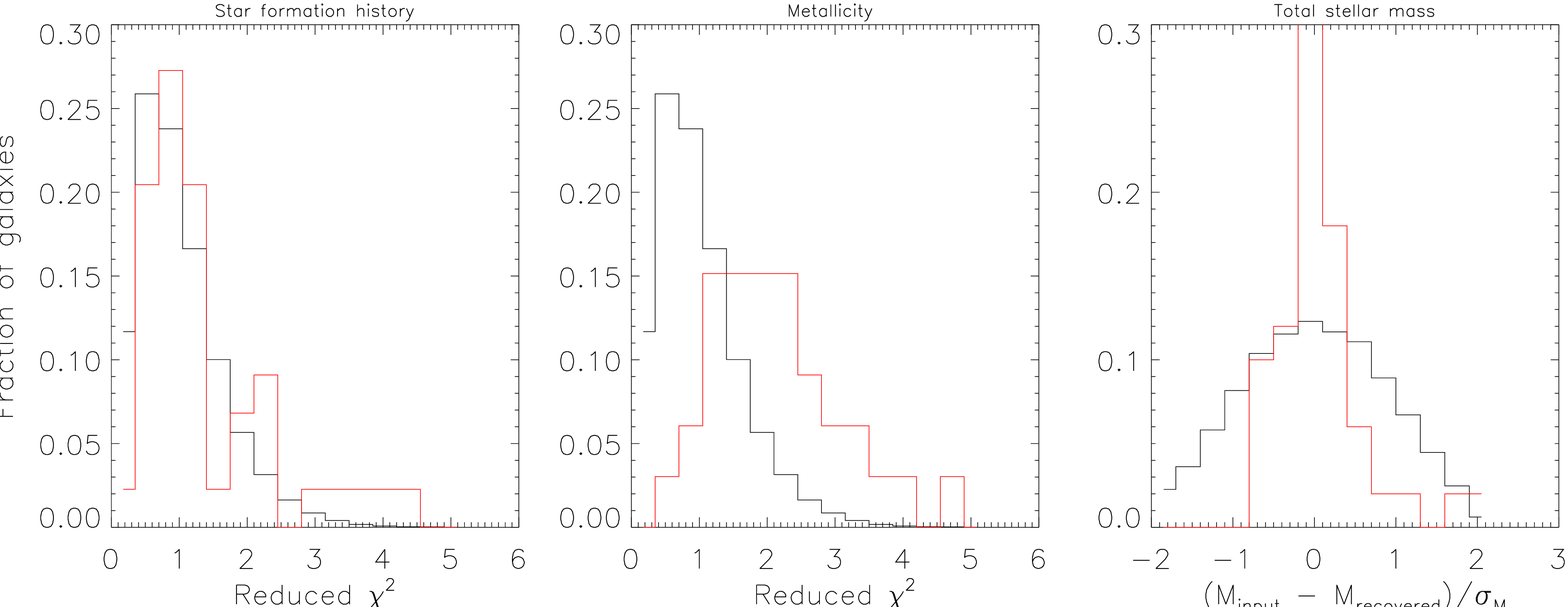}
\vspace{0.1in}
\caption{$\chi^2_{SFH}, \chi^2_{Z}$ and $(M^I - M^R)/\sigma_M$ for 50 mock galaxies with an exponentially decaying star formation history, at a redshift of 0.05 (red solid line). For reference, we also show the expected distributions in the same binning scheme (black solid line). The analysis is the same as seen in the red solid line of Figure \ref{fig:mocks_tau03_nerror50}, but with an extended wavelength coverage. See main text for discussion.}
\label{fig:mocks_tau03_nerror50_full}
\end{center}
\end{figure*}

As a test, we repeat the analysis done above, but this time using the full spectral range between  $[3800, 9200]$ \AA\ in the Earth-frame. We use the same galaxies as above, in the case of an exponentially decaying SFH, and at a redshift of 0.05. The results can be seen in Figure \ref{fig:mocks_tau03_nerror50_full}.  The red line in this Figure should be compared to the red solid line of Figure \ref{fig:mocks_tau03_nerror50} - the two distributions are very similar, indicating that removing these regions from the wavelength coverage does not affect the results in any significant way.

\section{Data} \label{sec:data}

We analysed the final data releases of the Main Galaxy Sample (MGS) and the Luminous Red Galaxies Sample (LRGS). \\

The main galaxy sample \citep{StraussEtAl02} is a magnitude-limited,
high-completeness ($>99\%$) galaxy sample, selected in the $r$-band. In its final data release it consists of over 700,000 galaxies, with a median redshift of approximately 0.11. Galaxies are selected according to three criteria: a star-galaxy
separation test, a cut in the $r'$-band petrosian magnitude (from here on we will refer to SDSS's $r'$- as $r$-band, and similarly to the other four bands), and a cut
in surface brightness. To avoid very poor quality data we impose a
further limit on the surface brightness $\mu < 23$.\\


The LRGs sample is selected based on $g-r$ and $r-i$ colors and $r$-band and $i$-band magnitudes, as detailed in \cite{EisensteinEtAl01}. The cuts were designed to follow an LRG fiducial model, and targets galaxies which are intrinsically bright and red from a redshift of around 0.2 to 0.6. \\

In both cases we use the bulk spectra samples provided in \\
http://www.sdss.org/dr7/products/spectra/getspectra.html

\subsection{Quality of the fits}
As in \cite{TojeiroEtAl07}, we find that the values of reduced $\chi^2$ still fall short of what is formally a good fit. The reason remains the limitations of the models in describing all the features of real spectra. This is visible in the plotted residuals of Figure \ref{fig:examples}, defined here as the difference between the input and the recovered spectrum, in units of the noise per wavelength bin. The left over structure on these residuals can in fact tell us something about what these limitations are, and with careful analysis tell us about which sets of models give a better prescription of real galaxies - see \cite{PanterEtAl07}. \\

The data reduction procedure can also impact on the quality of the fits. We find that the new spectro-photometric calibration pipeline, introduced in DR6, impacts visibly and positively on the quality of the fits we recover (see Section \ref{sec:masses_dr7}), but still does not allow values of reduced $\chi^2$ of unity. For further discussion, refer to \cite{TojeiroEtAl07}.

\subsection{Handling SDSS data}\label{sec:handling_data}

\subsubsection{Galactic extinction}
Spectra in DR7 do not include a correction for dust extinction
due to our own galaxy. We use the Galactic dust maps by \cite{SchlegelEtAl98} 
to obtain a value of E(B-V) for each spectroscopic plate. We estimate of the un-obscured flux using the dust extinction curve
of \cite{OdonnellEtAl94}, which assumes a uniform dust screen.  

\subsubsection{Pre-processing}
Prior to any analysis, we processed the SDSS spectroscopic data, so as
to accomplish the desired spectral resolution and
mask out any unwanted signal.\\
\par\noindent
The SDSS data files supply a mask vector, which flags any potential
problems with the measured signal on a pixel-by-pixel basis. We use
this mask to remove any unwanted regions and emission lines. In
practical terms, we ignore any pixel for which the provided mask value
is not zero. \\
\par\noindent
The BC03 synthetic models produce outputs at a resolution of 3\AA,
which we convolve with a Gaussian velocity dispersion curve with a
stellar velocity $\sigma_V= 170$km$s^{-1}$, this being a typical value
for SDSS galaxies \citep{PanterEtAl07}. \cite{PanterEtAl07} have also shown that there is no significant effect on the recovered star formation and metallicity histories as a result of the adoption of a single value of velocity dispersion. At the expense of CPU time, one can lift this limitation, but the data is currently too poor to justify this. One may worry about a metallicity - velocity dispersion degeneracy, as reported in \cite{KolevaEtAl08}. This is at the moment of no concern, given the quality of the data and the dependence of the recovered metallicity values on the SSP model chosen (see Section \ref{sec:metallicity}), which is dominant. The M05 models are given at a resolution of 20\AA, and we do not apply any further dispersion in this case.
We take the models' tabulated wavelength values as a fixed grid and
re-bin the SDSS data into this grid, using an inverse-variance weighted
average. We compute the new error vector accordingly. Note that the
number of SDSS data points averaged into any new bin is not constant,
and that the re-binning process is done after we have masked out any
unwanted pixels. Additionally to the lines yielded by the mask vector,
we mask out the following emission line regions in every spectrum's rest-frame
wavelength range: [5885-5900,
  6702-6732, 6716-6746, 6548-6578, 6535-6565, 6569-6599, 4944-4974,
  4992-5022, 4846-4876, 4325-4355, 4087-4117, 3711-3741, 7800-11000]
\AA. These regions were determined by visual inspection of over 1000
galaxy fit residuals \citep{PanterEtAl07}. \\
\par\noindent
These re-binned data- and noise-vectors are essentially the ones we use in our
analysis. However, since the linear algebra assumes white-noise, we pre-whiten the data and construct a new
flux vector $F'_{j} = F_{j}/\sigma_j$, which has unit variance, $\sigma'_j = 1, \forall
j$, and a new model matrix $A'_{ij} = A_{ij}/\sigma_j$.\\

The wavelength vector is shifted to the galaxy's rest frame. 

\section{The catalogue} \label{sec:catalogue}

The catalogue is published as a query-based T-SQL database. The data is
organised in tables, which we describe in section \ref{sec:database}. The principal
physical properties provided by VESPA are summarised in Table
\ref{tab:VESPA_out} .
\begin{table}[htdp]
\begin{center}
\begin{tabular}{|c|c|p{2in}|}
\hline
Symbol & Units & Description \\ \hline
\hline
$x_\alpha$ & - & star formation fractions in bin $\alpha$ \\  \hline
$u_\alpha$ & $M_\odot$ & stellar mass formed in bin $\alpha$ \\ \hline
$m_\alpha$ & $M_\odot$ & recycled stellar mass in bin $\alpha$ \\ \hline
$C_{\alpha\beta}$ & $M_\odot^2$ &covariance matrix for the star formation fractions \\ \hline
$Z_\alpha$ &  - & mass-weighted metallicity for bin $\alpha$ \\ \hline
$C^Z_{\alpha\beta}$ & - & covariance matrix for the metallicities \\ \hline
$M_*$ & $M_\odot$ & recycled stellar mass in galaxy \\ \hline
$M^u_*$ & $M_\odot$ & total stellar mass formed in galaxy \\ \hline
$\tau_V^{ISM}$ & - & dust extinction due to the inter-stellar medium for all populations \\ \hline
$\tau_V^{BC}$ & - &  dust extinction due to the birth cloud for young populations \\
\hline
\end{tabular}
\end{center}
\caption{Galaxy properties which are derived by VESPA. Where appropriate, quantities are corrected for fiber aperture.}
\label{tab:VESPA_out}
\end{table}%



In the next section we describe the meaning and calculation of each of these quantities.

\subsection{Masses and mass fractions}
\label{sec:cat_massfractions}
VESPA recovered star formation fractions, which we transform into absolute
masses as follows.  \\

If $\lambda$ is in the observed frame, then we can relate flux of a galaxy to the emitted luminosity by
\begin{equation}
F_\lambda = \frac{L_{\lambda / (1+z)}}{4\pi D_L^2 (1+z)}
\label{eq:flux_luminosity}
\end{equation}


In practical terms we do the following. We can rewrite equation (\ref{eq:vespa_problem}) in a simplified way:

\begin{equation}
F_j = \sum_\alpha x_\alpha S_{\alpha j}
\end{equation}

where $F_j$ has been shifted to the rest frame of the source, and $x_\alpha$ is the star formation fraction formed in $\Delta t_\alpha$.


The luminosity of a galaxy, written in terms of our models, is simply
$L_j = \sum_\alpha u_\alpha S_{\alpha j}$, where $u_\alpha$ is the
stellar mass formed at age $\alpha$. From (\ref{eq:flux_luminosity}): 

\begin{equation}
F_j = \frac{\sum_\alpha u_\alpha G_{\alpha j}} {4\pi D_L^2 (1+z)},
\label{eq:flux}
\end{equation}

which for any given age bin $\alpha$ gives the mass formed in each bin $\alpha$ in
units of solar masses: 

\begin{equation}
\label{eq:x_to_u}
u_\alpha = x_\alpha 4\pi D_L^2(1+z).
\end{equation}

\par\noindent
We distinguish between the stellar mass ever formed in a galaxy, and
the stellar mass remaining in a galaxy today: 

\begin{equation}
\label{eq:mass_t}
M(t) = \int_0^t \psi(t') \left[ 1-R(t-t')\right] dt'
\end{equation}

where $R(t-t')$ is the fraction of stellar mass lost to the ISM at
time $t$, by a stellar population aged $t'$ and $\psi(t')$ is the star
formation rate at age $t'$. In practical terms we calculate the
following: 

\begin{equation}
m_\alpha = u_\alpha  R_\alpha
\end{equation}
\begin{equation}
M_{*, fiber}^u = \sum_\alpha u_\alpha
\end{equation}
\begin{equation}
M_{*, fiber} = \sum_\alpha m_\alpha
\end{equation}

where $R_\alpha$, known as the {\it recycling fraction} is given by the models, for each of the metallicities. $R_\alpha$ is typically of the order of $0.5$ for the older populations, whereas in the younger bins the mass loss is much less significant with $R_\alpha$ between $0.7$ and $0.9$, depending on the width of the bin.\\

Finally, we correct for the fact that the fiber has an aperture of 3
arcseconds, which means we do not typically observe the entirety of
the galaxy. We use the observed fiber and petrosian magnitudes in the
$z$-band to scale up the stellar mass as 

\begin{equation}
\label{eq:scaled_mass}
M_* = M_{*, fiber} \times 10^{0.4(z_p - fz_p)}
\end{equation}

where $z_p$ and $fz_p$ are the petrosian and fiber magnitudes in the
$z$-band respectively. This scaling assumes that the parts of the
galaxy which do not fall under the fiber's aperture have an identical
star formation history as that observed. For an ensemble of galaxies fiber aperture
corrections are not important, in the sense that the mean color from
the fiber is the same as the mean color from the photometry \citep{GlazebrookEtAl03}. However,
one should keep in mind that they remain important for individual
galaxies, or at very low redshift. 

\subsection{Error estimates} \label{sec:errors}

To estimate how much noise affects our recovered solutions we take a rather
empirical approach. For each recovered solution we create $n_{error}$
random noisy realisations and we apply VESPA to each of these
spectra. In the current runs we have used $n_{error}=20$. We re-bin each recovered solution in the parametrization of
the solution we want to analyse and estimate the
covariance matrices
\begin{equation}
\label{eq:cx}
C(x)_{\alpha \beta} = \left\langle (x_\alpha - \bar{x}_\alpha)(x_\beta - \bar{x}_\beta)
\right\rangle
\end{equation}
\begin{equation}
\label{eq:cz}
C(Z)_{\alpha \beta} = \left\langle (Z_\alpha - \bar{Z}_\alpha)(Z_\beta - \bar{Z}_\beta) \right\rangle.
\end{equation}

For convenience, we also
define a covariance matrix of the unrecycled mass per bin as (using
equation \ref{eq:x_to_u})

\begin{equation}
\label{eq:Cu}
C_{\alpha\beta}(u) = C_{\alpha\beta}(x) \left[ 4\pi D_L^2(1+z)\right] ^2
\end{equation}

from which we can estimate the error in the unscaled mass formed in
bin $\alpha$ as $\sigma_{u(\alpha)} =
\sqrt{C_{\alpha\alpha}(u)}$. This ignores the uncertainty in the
estimation of the redshift, which is of little significance compared
to the variance of $u_\alpha$ across realisations. From
$\sigma_{u(\alpha)}$ we can calculate errors in $\sigma_{m(\alpha)}$
by multiplying by the corresponding recycling fraction
$R_\alpha$. The error on the metallicity is simply
$\sqrt{C_{\alpha\alpha}(Z)}$. \\

We use the full covariance matrix to estimate the statistical error on
the total stellar mass, $M_*$:

\begin{equation}
\label{eq:error_mass}
\sigma^2(M_*) = \sigma^2(M_{*,fiber}) \gamma^2 + M^2_{*,fiber} \sigma^2(\gamma)
\end{equation}

where $\gamma$ is the conversion factor between fiber and galaxy mass
of equation (\ref{eq:scaled_mass}), and $\sigma(\gamma)$ is the error
associated with this factor, calculated using the errors in the petrosian and fiber
$z$-band magnitudes. $\sigma^2(M_{*,fiber})$ is estimated from the
unrecycled mass covariance matrix and the total recycling fraction
of the galaxy, $R$:

\begin{equation}
\sigma^2(M_{*,fiber}) = \sum_{\alpha, \beta} C_{\alpha \beta}(u)R^2.
\end{equation}

This assumes that there is no error in the recycling fractions - i.e.,
that we know the SFH exactly. \\

This approach estimates errors due to photon noise only, but there is also a systematic error associated with the limitation of the models we use. The true effect of the models on the recovered solutions is impossible to calculate. By providing solutions using more than one set of models we give the user an opportunity to check how the answers of interest to them change by using different models. However, we caution against using this as a quantitative estimate of the systematic error associated with modeling, as models could be wrong or incomplete in the same way.\\

\begin{figure}[htbp]
\begin{center}
\epsscale{1}
\plotone{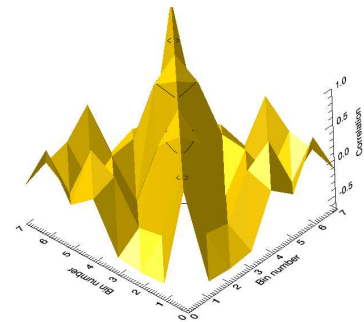}
\caption{The correlation matrix of the recovered mass fractions for the galaxy in the left-hand side panel of Figure \ref{fig:examples}.}
\label{fig:r}
\end{center}
\end{figure}

As an illustration of the type of correlations one finds between the different age bin, in Figure \ref{fig:r} we show the correlation matrix for the galaxy in the left-hand panel of Figure \ref{fig:examples}. The correlation function is defined as
\begin{equation}
r_{\alpha,\beta} = \frac{C_{\alpha,\beta}(u)}{\sigma_\alpha \sigma_\beta}
\end{equation}

and shows how independent two quantities are - in this case $m_\alpha$ and $m_\beta$. By construction $r_{\alpha,\beta} \in [-1,1]$. One can see that the highest cases of correlation happen in adjacent bins, which is not surprising as the spectral signature of two adjacent bins is more similar than that of two bins further apart in lookback time. Non-adjacent bins however, are not completely uncorrelated, with the highest absolute of value of $r$ being around $0.6$.

\subsection{Mass and metallicity per age bin} \label{sec:bins}

The history of each individual galaxy is likely to be parametrized by a
combination of high- and low-resolution age bins. One must be careful
to interpret the masses and metallicities associated with
low-resolution bins. We re-iterate that in these cases the mass recovered
should be interpreted as the total mass recovered in the bin, but we
have little information of when, within the bin, it formed. Similarly, we
should interpret the metallicity values recovered as a mass-weighted
metallicity for the whole duration of the bin. This information can be accessed via the table BinProp, see Section \ref{sec:database} and Table \ref{tab:BinProp}.\\

In addition to the masses and metallicities obtained in their original resolution, we choose to also provide a fully-resolved star formation and metallicity history for each galaxy. In this case, the solutions are {\it post-processed} to be presented in 16 bins. We do this by using the weights of equation (\ref{eq:weights}) to split the mass of a low-resolution bin across the relevant high-resolution bins. We use the same weights for the error in the mass. The metallicity of the new high-resolution bins will be set to be the same as the old low-resolution bins, and the error remains the same. This information can be accessed via the table HRBinProp, see section \ref{sec:database} and Table \ref{tab:HRBinProp}.\\

By construction, the solutions in HRBinProp are always consistent with the ones in binProp, but they should be used with caution. HRBinProp is a very useful table when one needs to have a uniform set of bins in order to study, for example, average rest-frame quantities as a function of time. Using this table for ensemble of galaxies is appropriate, but not on an individual galaxy basis. Care must also be taken with the treatment of mass errors across the high-resolution bins, as they will of course be highly correlated with those bins which formed the same low-resolution bin. When taking averages over large ensembles, an error on the mean of each individual bin might be more appropriate. \\

\subsection{Dust}

Depending on the dust model used we recover either one or two values of dust extinction, one associated with young stars
($\tau_V^{BC}$ - applied to stellar populations younger than
0.03 Gyrs) and one associated with the whole galaxy
($\tau_V^{ISM}$ - applied to all stellar populations), following the
mixed slab model of \cite{CharlotFall00}. \\

The two-parameter dust model gives a more realistic description of the data, but it is not clear whether the data justifies the extra degree of freedom. In \cite{TojeiroEtAl07} we saw that even if $\tau_V^{BC}$ is slightly degenerate with the amount of stellar mass created in populations younger than $t_{BC}$, it helps to recover a more accurate value of $\tau_V^{ISM}$. The solution with a one-parameter model is a necessary step to a solution with a two-parameter model (see \cite{TojeiroEtAl07}), and we see no reason to not make the two sets of solutions public, especially if the user is interested in populations younger than $t_{BC}$. However, we do not run a full error-analysis on the one-parameter solution, and all the error columns are set to zero. 

\section{The database}
\label{sec:database}
The catalogue is being published as a relational T-SQL database, which provides
maximum flexibility and interoperability with other databases. It can be accessed through the WFCAM Science Archive (WSA) via http://www-wfau.roe.ac.uk/vespa/ and will be made
available on Astrogrid in due course. Information about subsequent data releases using new models, code versions or data-sets will be given on this website, and any information online overrides information on this paper.\\

The database is split
into a number of tables, each with a number of fields (columns), which
can be accessed and queried via SQL.\\

We have split the database into seven tables: 
\begin{itemize}
\item GalProp includes results
relative to the galaxy as a whole, 
\item BinProp has results which refer to
each bin individually, in the original resolution as recovered by VESPA
\item HRBinProp splits the recovered solutions into the highest possible resolution (16 bins), using the appropriate weights which are given by equation (\ref{eq:weights}).
\item DustProp holds dust information 
\item BinID identifies each VESPA low- or high-resolution bin. 
\item RunProp holds run-specific details, such as SSP and dust models used.
\item LookupTable links VESPA's unique identifier with SDSS's own identifiers, such as specObjID, plate, MJD and fiber ID.
\end{itemize}

Tables \ref{tab:GalProp}, \ref{tab:BinProp}, \ref{tab:HRBinProp}, \ref{tab:DustProp} and
\ref{tab:BinID} detail each of the fields included in the
tables mentioned above. Galaxies are identified by a unique index, which can be associated with 
SDSS's specObjID via the table LookupTable. This means that object properties which are already
included in the SDSS need not be included in this database, as
cross-matching can be done in a straightforward way using specObjID. \\

\begin{table}[htdp]
\begin{center}
\begin{tabular}{|c|c|p{2in}|}
\hline
Field & Units & Description \\ \hline
\hline
indexP & & Unique identifier, constructed from SDSS's plate, MJD and fiberID info.\\ \hline
runID & & Gives detail of the run, in RunProp. \\ \hline
m\_stellar & $M_\odot$ & $M_*$  - equation (\ref{eq:scaled_mass}).\\ \hline
m\_stellar\_error & $M_\odot$ & $\sigma(M_*)$ - equation (\ref{eq:error_mass}).\\ \hline
t\_lb & Gyr & Lookback time of galaxy, assuming a WMAP5 cosmology. \\ \hline
chi2 & & $\chi^2$ of the unmasked regions used for spectral fit. \\ \hline
SNR & & Signal to noise ratio of the used (unmasked) spectrum, at the models' resolution. \\ \hline
nbins & & Number of recovered bins in a galaxy. \\ \hline
npops & & Number of recovered bins with non-zero mass. \\ \hline
\end{tabular}
\end{center}
\caption{\label{tab:GalProp}GalProp}
\end{table}%

\begin{table}[htdp]
\begin{center}
\begin{tabular}{|c|c|p{2in}|}
\hline
Field & Units & Description \\ \hline
\hline
indexP & & Unique identifier, constructed from SDSS's plate, MJD and fiberID info.\\ \hline
runID & & Gives detail of the run, in RunProp. \\ \hline
binID & & Bin identifier as given by Figure \ref{fig:bins_table}. \\ \hline
mass & $M_\odot$ & Mass formed in bin - equation (\ref{eq:x_to_u}) - corrected for fiber aperture. \\ \hline
mass\_error & $M_\odot$ & $\sigma_{u(\alpha)}$ - as derived from equation (\ref{eq:Cu}) - corrected for fiber aperture.\\ \hline
SFR & $M_\odot$Gyr$^{-1}$ &Star formation rate in bin.\\ \hline
Z & & Metallicity in bin. \\ \hline
Z\_error & & $\sigma_{Z(\alpha)}$ as derived from equation (\ref{eq:cz}). \\ \hline
\end{tabular}
\end{center}
\caption{BinProp}
\label{tab:BinProp}
\end{table}%

\begin{table}[htdp]
\begin{center}
\begin{tabular}{|c|c|p{2in}|}
\hline
Field & Units & Description \\ \hline
\hline
indexP & & Unique identifier, constructed from SDSS's plate, MJD and fiberID info.\\ \hline
runID & & Gives detail of the run, in RunProp. \\ \hline
binID & & Bin identifier as given by Figure \ref{fig:bins_table}. \\ \hline
mass & $M_\odot$ & Mass formed in bin - from equation (\ref{eq:x_to_u}) with weights from equation (\ref{eq:weights}) and corrected for fiber aperture. \\ \hline
mass\_error & $M_\odot$ & $\sigma_{u(\alpha)}$ - from equation (\ref{eq:Cu}) with weights from equation (\ref{eq:weights}) and corrected for fiber aperture.\\ \hline
Z & & Metallicity in bin. \\ \hline
Z\_error & & $\sigma_{Z(\alpha)}$ as derived from equation (\ref{eq:cz}). \\ \hline
\end{tabular}
\end{center}
\caption{HRBinProp}
\label{tab:HRBinProp}
\end{table}%

\begin{table}[htdp]
\begin{center}
\begin{tabular}{|c|c|p{2in}|}
\hline
Field & Units & Description \\ \hline
\hline
indexP & & Unique identifier, constructed from SDSS's plate, MJD and fiberID info.\\ \hline
runID & & Gives detail of the run, in RunProp. \\ \hline
dustID & & Dust identifier: 1 for $\tau_V^{BC}$ and 2 for $\tau_V^{ISM}$. \\ \hline
dustVal & & Either $\tau_V^{BC}$ or $\tau_V^{ISM}$, according to dustID. \\ \hline
\end{tabular}
\end{center}
\caption{DustProp}
\label{tab:DustProp}
\end{table}%

\begin{table}[htdp]
\begin{center}
\begin{tabular}{|c|c|p{2in}|}
\hline
Field & Units & Description \\ \hline
\hline
binID & & Bin identifier, as given by Figure \ref{fig:bins_table}. \\ \hline
ageStart & Gyr & Age of the young boundary of the bin. \\ \hline
ageEnd & Gyr & Age of the old boundary of the bin. \\ \hline
width & Gyr & Width of the bin in Gyrs. \\ \hline
width\_bin & & Width of the bin in units of high-resolution bins. \\ \hline
\hline
\end{tabular}
\end{center}
\caption{BinID}
\label{tab:BinID}
\end{table}%

\begin{table}[htdp]
\begin{center}
\begin{tabular}{|c|c|l|}
\hline
Field & Units & Description \\ \hline
\hline
runPropID & & Unique run ID. \\
SSP & & SSP models used. \\
codeVersion & & VESPA code version. \\
dustModel & & Dust model used. \\
dataRelease & & SDSS's data release. \\
sample & & MGS or LRG. \\
\hline
\end{tabular}
\end{center}
\caption{RunProp}
\label{tab:RunProp}
\end{table}%

Table \ref{tab:RunProp_full} lists the details of the VESPA runs which are currently complete, and are either on the database at the time of publications or will be in a very short period. Future extensions of the catalogue or code will increase this table accordingly. \\

\begin{table}[htdp]
\begin{center}
\begin{tabular}{|c|c|c|c|c|c|}
\hline
runID & SSP & codeVersion & dustModel & dataRelease & sample \\ \tableline

1 & BC03 & 1.0 & 1 & DR7 & MGS \\
2 & BC03 & 1.0 & 2 & DR7 & MGS \\
3 & M05 & 1.0 & 1 & DR7 & MGS \\
4 & M05 & 1.0 & 2 & DR7 & MGS \\
5 & BC03 & 1.0 & 1 & DR7 & LRG \\
6 & BC03 & 1.0 & 2 & DR7 & LRG\\
7 & M05 & 1.0 & 1 & DR7 & LRG \\
8 & M05 & 1.0 & 2 & DR7 & LRG \\
\hline
\end{tabular}
\end{center}
\caption{RunProp in full}
\label{tab:RunProp_full}
\end{table}%

\subsection{Example queries}

In this section we give some specific examples of how to explore the catalogue. This list is simply intended to demonstrate some of the potential of the catalogue, and give the user a starting point.\\

A simple query to return the present-day stellar mass of all galaxies in the LRG sample, analysed with BC03 and a 2-parameter dust model could look like

\begin{verbatim}
SELECT specObjID, m_stellar, m_stellar_error
FROM lookUpTable as l, galProp as g
WHERE
g.indexP = l.indexP
AND g.runID = 6
\end{verbatim}

The rest-frame averaged star formation fractions as a function of lookback time, for the MGS galaxies, analysed with M05 and a 2-parameter dust model is accessible via the query

\begin{verbatim}

SELECT binID, AVG(hr.mass/g.m_stellar) 
FROM hrBinProp as hr, galProp as g 
WHERE 
g.indexP = hr.indexP 
AND g.runID=4
AND hr.runID=4
GROUP BY hr.binID 
ORDER BY binID
\end{verbatim}

It is also easy to select galaxies according to their star formation histories. For example, the following query returns the stellar mass distribution of galaxies which form over than 50\% of their present-day stellar mass in the last age bin:

\begin{verbatim}
SELECT 
.5+FLOOR(10.*LOG10(g.m_stellar))/10. as lgm, 
COUNT(*) as freq
FROM GalProp as g, BinProp as b,
binID as bID
WHERE
g.indexp=b.indexp 
AND bid.binid = b.binID
AND bid.ageEnd = 14
AND b.mass > (0.5 * g.m_stellar)
AND g.m_stellar < 1e14
AND g.m_stellar > 1e7
AND g.runID = 2
AND b.runID = 2
GROUP BY .5+FLOOR(10.*LOG10(g.m_stellar))/10.
ORDER BY lgm
\end{verbatim}

We can also access average star formation rates over any period of time, by targeting the correct bins. To return the average star formation rate over the last 115 Myrs of the lifetime of the galaxy (corresponding to the first five high-resolution bins), one can do

\begin{verbatim}
SELECT specObjID, SUM(hr.mass)/115. as SFR
FROM
lookUpTable as l, hrBinProp as hr, 
binID as bID
WHERE
l.indexP = hr.indexP
AND hr.binID = bID.binID
AND hr.binID < 5
AND hr.runID = 4
GROUP BY specObjID
\end{verbatim}

To access the star formation history of any given galaxy the user can select, for example, on plate, fiberID and MJD as:

\begin{verbatim}
SELECT specObjID, bID.ageStart, bID.ageEnd, 
b.mass, b.mass_error, b.Z, b.Z_error
FROM
binID as bID, binProp as b, 
lookUpTable as l
WHERE
l.indexP = b.indexP
AND b.binID = bID.binID
AND l.plate = 0353
AND l.mjd = 51703
AND l.fiberID = 077
AND runID = 2
\end{verbatim}

We re-iterate that when selection from binProp the number of rows returned will vary from one galaxy to the next. If a uniform number of bins is required, then use HRBinProp but this should be done with caution (see Section \ref{sec:bins}). The user should also be careful to always make a selection on runID {\it on all relevant tables} to avoid duplicate results - each galaxy might be represented in the catalogue a number of times, analysed with a different combination of models. To run the same query with any different set of models, the user needs only to change runID. \\

Whereas these queries are focused on rest-frame quantities, we also provide the redshift and look-back time of each galaxy in the table GalProp, for easy access to Earth-frame quantities.\\

It is also possible to directly tap into the SDSS DR7 databases that are held at the WSA. For example, return fiber-corrected stellar mass and the fiber-corrections which were applied to those masses one can do:

\begin{verbatim}
SELECT l.specObjID, g.m_stellar, 
sdss.petromag_z - sdss.fibermag_z AS fiber_diff, 
power(10.0000,0.4*(sdss.petromag_z - 
sdss.fibermag_z)) 
AS fiber_corr
FROM bestDR7..specPhotoAll as SDSS, 
galprop as g, lookuptable as l
WHERE
sdss.specObjID = l.specObjID
AND l.indexp = g.indexp
AND g.runid=2
\end{verbatim}



Details on the other databases at the WSA can be found at http://surveys.roe.ac.uk/wsa/.

\section{Results} \label{sec:results}
The VESPA catalogue can be exploited in many and varied ways. In here, we take the opportunity to show some basic characteristics of the data. We explore average rest-frame star-formation histories, stellar masses, dust content and modeling, star formation rates and show some comparisons with GALEX magnitudes. \\

We will mention red and blue galaxies throughout, which refer to a simple cut in $u-r$ color. We have assigned galaxies with $u-r < 1.5$ the index of blue and to those with $u-r > 2.8$ the index of red. 

\subsection{Number of parameters}
\begin{figure}
\plotone{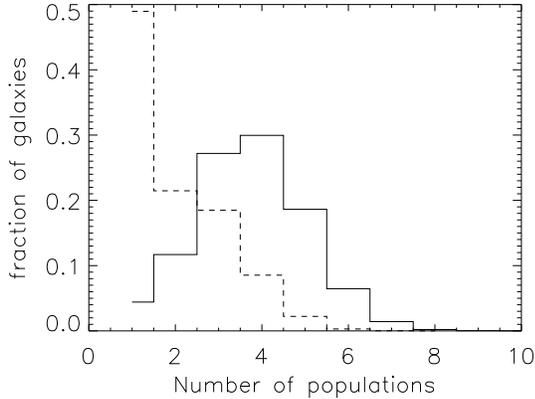}
\caption{Number of populated mass bins for the two samples of galaxies currently in the catalogue: the MGS (in the solid line) and the LRG sample (in the dashed line). }
\label{fig:npops}
\end{figure}

First let us look at the typical number of non-zero mass bins recovered for each galaxy, which is shown in Figure \ref{fig:npops}. As first shown in \cite{TojeiroEtAl07}, VESPA recovers typically between two to five populations from each galaxy in the MGS. This histogram is dramatically different for the LRG sample, with the majority of the galaxies being parametrized with three or less populations. This is likely to be a combination of two factors: LRGs have simple star formation histories, dominated by old stars; and the spectral data for LRGs is of poorer quality, given the larger redshift range of this sample. This suggests that parametrizing the LRGs with a single burst at high-redshift may be an acceptable solution, especially given the typical quality of the data. However, such a simplification is not justifiable for a MGS galaxy, and the SDSS data certainly allows for a much better description. 

\subsection{Stellar masses in DR7}\label{sec:masses_dr7}
\begin{figure}
\plotone{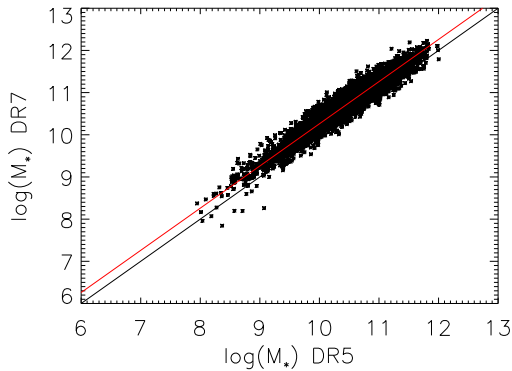}
\caption{Stellar masses from a random sub-sample of galaxies analysed using their DR5 and DR7 spectra. The difference is due to a new flux scale introduced in DR6 \citep{Adelman-McCarthyEtAl08}. The constant offset which best fits the data is 0.26 dex, and is represented by the red line. }
\label{fig:dr5_dr7_masses}
\end{figure}

SDSS's DR6 introduced a change in the spectroscopic calibration scale of approximately 0.35 magnitudes - spectra from DR6 and onwards are {\it brighter} \citep{Adelman-McCarthyEtAl08}. The result is a nearly constant offset in the estimated masses between data releases before and after DR6, as can be seen in Figure \ref{fig:dr5_dr7_masses}. Assuming a slope of unity, we find that the offset which brings the two estimates into agreement is 0.26 dex. We find that the difference in other derived quantities, such as star formation histories and dust extinction is negligible. \\

\begin{figure}
\plotone{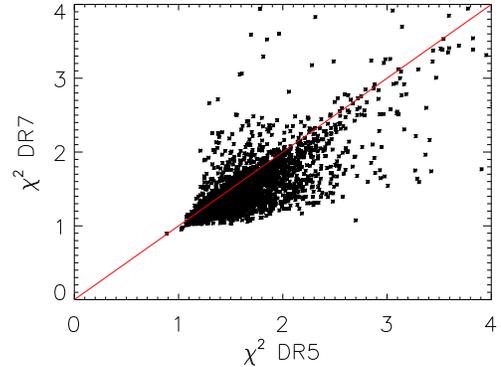}
\caption{$\chi^2$ of a random sub-sample of galaxies analysed using their DR5 and DR7 spectra. There is a visible improvement in DR7, in the majority of the cases.}
\label{fig:dr5_dr7_chi2}
\end{figure}

Figure \ref{fig:dr5_dr7_chi2} shows the difference in the $\chi^2$ values of the fits. We see that in the majority of cases the new spectro-photometric calibrations introduced with DR6 have somewhat improved the goodness-of-fit of the VESPA solutions.\\

\begin{figure}
\plotone{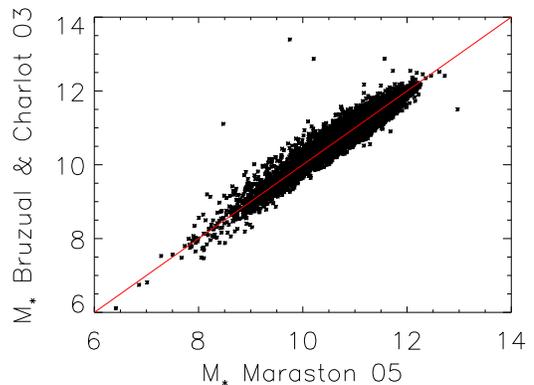}
\caption{Present-day stellar mass estimates, recovered using the BC03 and the M05 models, and a two-parameter dust model.}
\label{fig:masses_m05_bc03}
\end{figure}

The effect of the choice of SSP models on the estimation stellar mass of high-redshift galaxies is known to be significant \citep{MarastonEtAl06}. However, we find that under the SDSS's redshift and rest-frame wavelength range, estimates of stellar mass from full-spectral fitting are robust against the choice of SSP modeling as can be seen in Figure \ref{fig:masses_m05_bc03}. BC03 masses are only very slightly systematically higher but within the 1$\sigma$ errors. The reduced $\chi^2$ of the one-to-one line, taking into account the errors on both measurements is 1.3. This indicates that there is a source of error associated with the choice of SSP which is not accounted for in the formal errors returned by VESPA. This is expected, and we brought the attention to this matter in Section \ref{sec:errors}. However, there is no statistically significant offset. M05 models generally bring masses down due to the inclusion of the thermally pulsating asymptotic giant branch stars, which are very bright and generally require less stellar mass to explain a given luminosity (see also Section \ref{sec:average_sff}) but these stars have the most effect at wavelengths redwards of the SDSS spectral range, as seen in \cite{MarastonEtAl06}.\\

\subsection{Metallicity estimates}\label{sec:metallicity}

\begin{figure}
\plotone{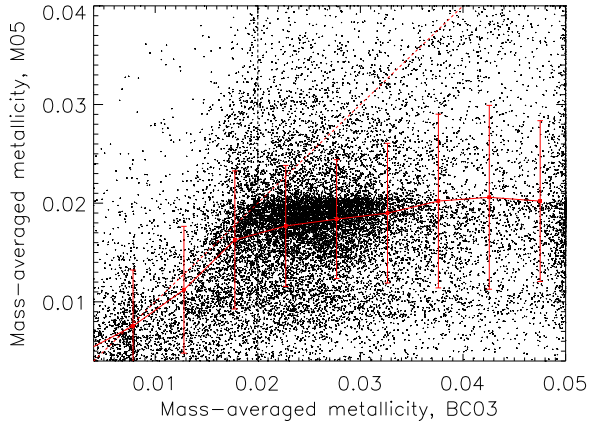}
\vspace{0.1in}
\caption{Mass-averaged metallicity for a random selection of 20,000 galaxies analysed with the BC03 and M05 models. The solid red line shows the mean, and the dashed red line is the one-to-one line. For reference, the vertical black dashed line is at solar metallicity.}
\label{fig:met_comparison_M05_BC03}
\end{figure}

For each galaxy, we can also compute the mass-averaged metallicity using the recovered star-formation history. In contrast with estimates for the total stellar mass, we find in Figure \ref{fig:met_comparison_M05_BC03} that the estimated metallicity of each individual galaxy depends heavily on the SSP modelling used. \\

Some of this scatter is due to poorly constrained populations which may have some weight in terms of mass, but not in terms of light contribution (in, for example, galaxies with very young stellar populations). However, the overall trend is very clearly something which is followed by the vast majority of galaxies. We cannot say at this stage which model is more correct, and we are left simply to emphasise that the mass-averaged metallicity of galaxies has to be interpreted within the limitations of each set of models.



\subsection{Average star-formation histories} \label{sec:average_sff}

\begin{figure}
\plotone{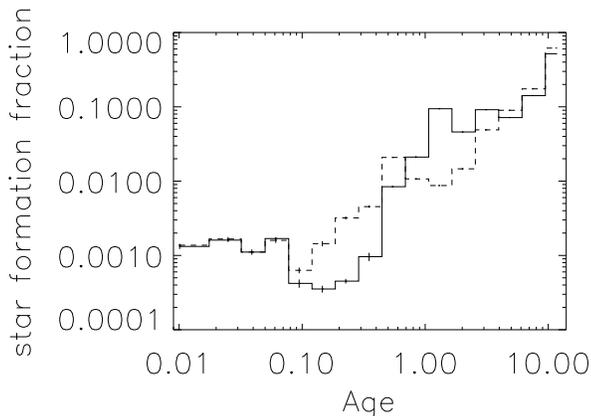}
\caption{The rest-frame averaged star formation history of a sub-sample of galaxies analysed using the BC03 models (solid line) and the M05 models (dashed line). The error bars are errors on the mean and we used a one-parameter dust model. }
\label{fig:average_sff_mgs}
\end{figure}

Figure \ref{fig:average_sff_mgs} shows the averaged rest-frame star formation history of a sample of 20,000 galaxies in the MGS, obtained using the BC03 and the M05 models. We do not intend to assess which sets of models are better, but rather identify where different models give very different results. In this case, for example, the inclusion of the thermally pulsating asymptotic giant branch stars in the M05 models means that populations of around 1Gyr are intrinsically brighter. This decreases the amount of mass needed in populations of this age, and results in a smoother decaying star formation history. \\

\begin{figure}
\plotone{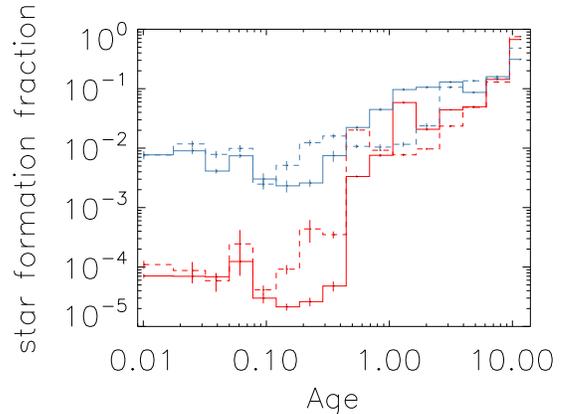}
\caption{The rest-frame averaged star formation history of a sub-sample of galaxies analysed using the BC03 models (solid lines) and the M05 models (dashed lines) for red and blue galaxies in red (lower) and blue (upper) respectively. The error bars are errors on the mean and we used a one-parameter dust model.}
\label{fig:average_sff_redblue}
\end{figure}

Figure \ref{fig:average_sff_redblue} shows the same for a subsample of blue and red galaxies. Whilst what we call red galaxies form most of their stars over 9 Gyrs ago, blue galaxies have a significantly flatter average star formation history and much more recent star formation, as expected. \\

\subsection{Star formation rates}
\begin{figure}
\plotone{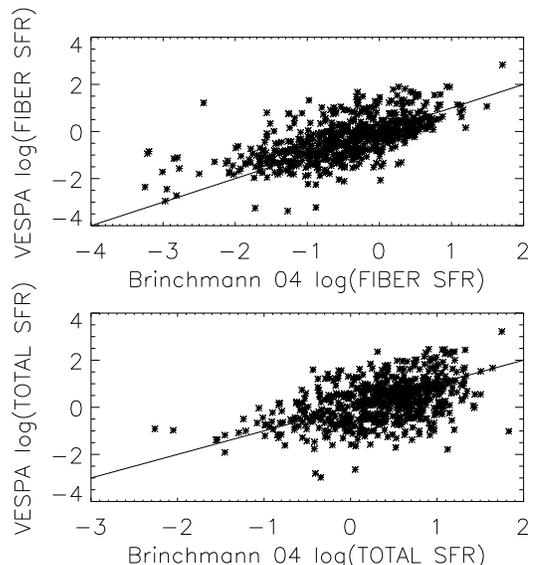}
\caption{The star formation rate estimated by averaging the mass formed over the latest 115 Myrs of each galaxy, using a two-parameter dust model and BC03 models. We compare these values to the SRF catalogue of \cite{BrinchmannEtAl04}, and first correct our recovered masses by 0.26 dex (see Section \ref{sec:masses_dr7}). The top panel shows the results within the 3-arcsecond fiber aperture, and the bottom panel the results extrapolated to the whole of the galaxy using equation (\ref{eq:scaled_mass}).}
\label{fig:SFR_comparisons}
\end{figure}

We can have a measurement of the instantaneous star formation rate by averaging the recent mass formed in a galaxy. We compared the SFR averaged over the last 115Myr - corresponding to the first 5 bins - to those of \cite{BrinchmannEtAl04}, which relate to DR4. Prior to this comparison we corrected our results for the mass offset of 0.26 dex, as discussed in the previous section. The results for both fiber and total quantities can be seen in Figure \ref{fig:SFR_comparisons}.\\

\subsection{Dust}

\begin{figure}
\plotone{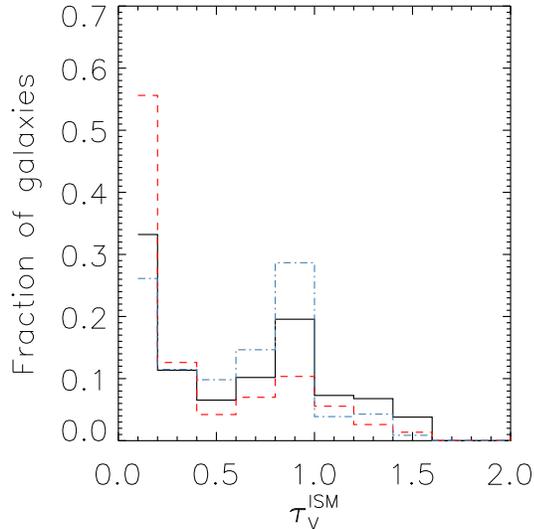}
\caption{Distribution of $\tau_V^{ISM}$ for a sample of galaxies, analysed with BC03 and a two-parameter dust model (black line). The red (dashed) and blue (dot-dashed) lines show the same but for sub-sample of red and blue galaxies, respectively. }
\label{fig:dust_histograms_BC03}
\end{figure}

VESPA provides an estimation of the dust content of a galaxy. Figure \ref{fig:dust_histograms_BC03} shows the recovered values of $\tau_V^{ISM}$ for a random subsample of approximately 10,000 galaxies analysed using the BC03 models, and a two-parameter model. We see a very bimodal distribution, which for red galaxies can be decoupled almost cleanly with our chosen color cut. This is very much what one would expect to see, with the red galaxies being more likely to populate the area of the histogram corresponding to a low dust attenuation. \\

\begin{figure}
\plotone{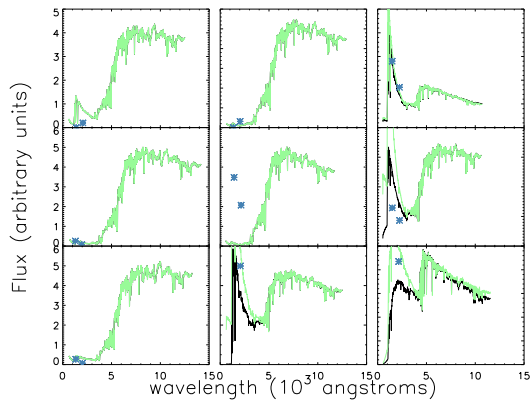}
\caption{The recovered spectrum of nine randomly selected galaxies, extrapolated into the UV. The black line is the recovered spectrum obtained with a two-parameter dust mode, and the over-plotted green line is the recovered spectrum obtained with a one-parameter dust model. The blue points are GALEX measured fluxes. For galaxies with very little or no recent star formation the two dust-models are in agreement, but differences are evident when young stars are present in the galaxy.}
\label{fig:UV_examples}
\end{figure}

The choice between a 1- or a 2-parameter dust model does not impact significantly on the recovered values of $\tau_V^{ISM}$, nor on the recovered star formation histories for the old populations. This happens because the young populations, which are the only ones affected by $\tau_V^{BC}$ are only significant, in terms of light, in the UV region of the spectrum. We exemplify this in Figure \ref{fig:UV_examples}, which shows the recovered spectrum of nine random galaxies extended to the UV. \\

\begin{figure}
\plotone{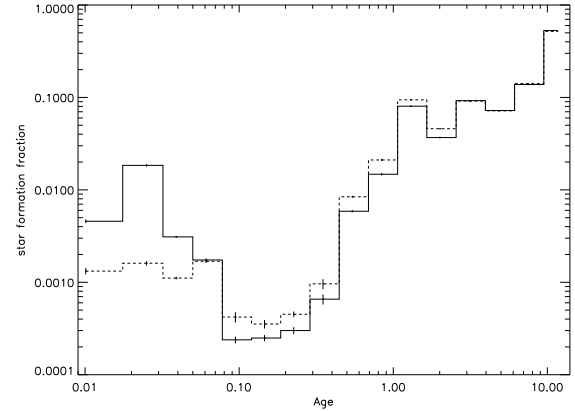}
\caption{The rest-frame averaged star formation history of a sample of galaxies analysed with a one-parameter dust model (dashed line) and a two-parameter dust model (solid line). The error bars are errors on the mean.}
\label{fig:average_sff_1p_2p}
\end{figure}

What is visibly affected by the choice of dust modeling however, is the amount of mass recovered in the first two bins, as expected. Figure \ref{fig:average_sff_1p_2p} shows the average rest-frame star-formation history of a sub-sample of MGS galaxies analysed with both a 1-parameter and a 2-parameter dust model, and the BC03 models. \\

In section \ref{sec:galex} we compare VESPA UV predictions with GALEX data to address the question of the need for the two-parameter dust model.

\subsubsection{Balmer decrements}
A direct comparison can be made with Balmer decrements between $H_\alpha$ and $H_\beta$ lines. Given that dust extinction is wavelength dependent, one can estimate the overall amount of extinction by studying the ratio of pairs of Balmer lines. If we know what the unextinguished Balmer decrement should be for a given pair, the expected Balmer decrement as a function of optical depth can be easily computed for the one-parameter dust model. We assumed $H_\alpha$/$H_\beta$ = 2.87 in the absence of dust \\

\begin{figure}
\plotone{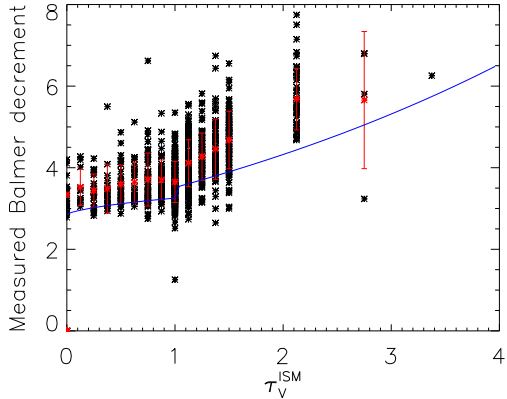}
\caption{The value of $\tau_V^{ISM}$ recovered by VESPA plotted against the measured Balmer decrement ($H_\alpha$/$H_\beta$) for a sample of galaxies analysed with a one-parameter dust model. The red stars are the mean value in each bin of $\tau_V^{ISM}$, and the error bars are errors on that mean. The blue line is the expected Balmer decrement for a one-parameter dust model as a function of $\tau_V^{ISM}$. The kink in the line at $\tau_V^{ISM}=1$ is a result from where we change from a mixed-slab to a uniform slab model (see Section \ref{sec:dust_modeling}).}
\label{fig:balmer_new}
\end{figure}

Using the DR7 release of the MPA added value catalogue \citep{KauffmannEtAl03aMNRAS, BrinchmannEtAl04}, we measured the Balmer decrement of a sample of galaxies, and compared it to the expected Balmer decrement given the recovered value of $\tau_V^{ISM}$. The result can be seen in Figure \ref{fig:balmer_new}. We recover the general behaviour as expected from the theory, but we measure values of the Balmer decrement which are systematically higher - although within the 1-sigma interval - given the dust content we recover with VESPA. We find the agreement encouraging. A slight adjustment of the expected value of the decrement in the absence of dust, combined with a steeper dust law (see also Section \ref{sec:galex}) would bring the two measurements closer. The galaxies for which a reliable Balmer decrement can be measured are typically star forming, and it is therefore not entirely surprising that $\tau_V^{ISM}$ is not a complete representation of the dust within the galaxy. Whereas Figure \ref{fig:balmer_new} suggests that quantitatively the actual values of optical depth one obtains using the two methods are offset, the good qualitative agreement allows the user to split any galaxy sample according to dust content with confidence.\\

\subsection{Comparisons with GALEX}\label{sec:galex}
\begin{figure*}
\plottwo{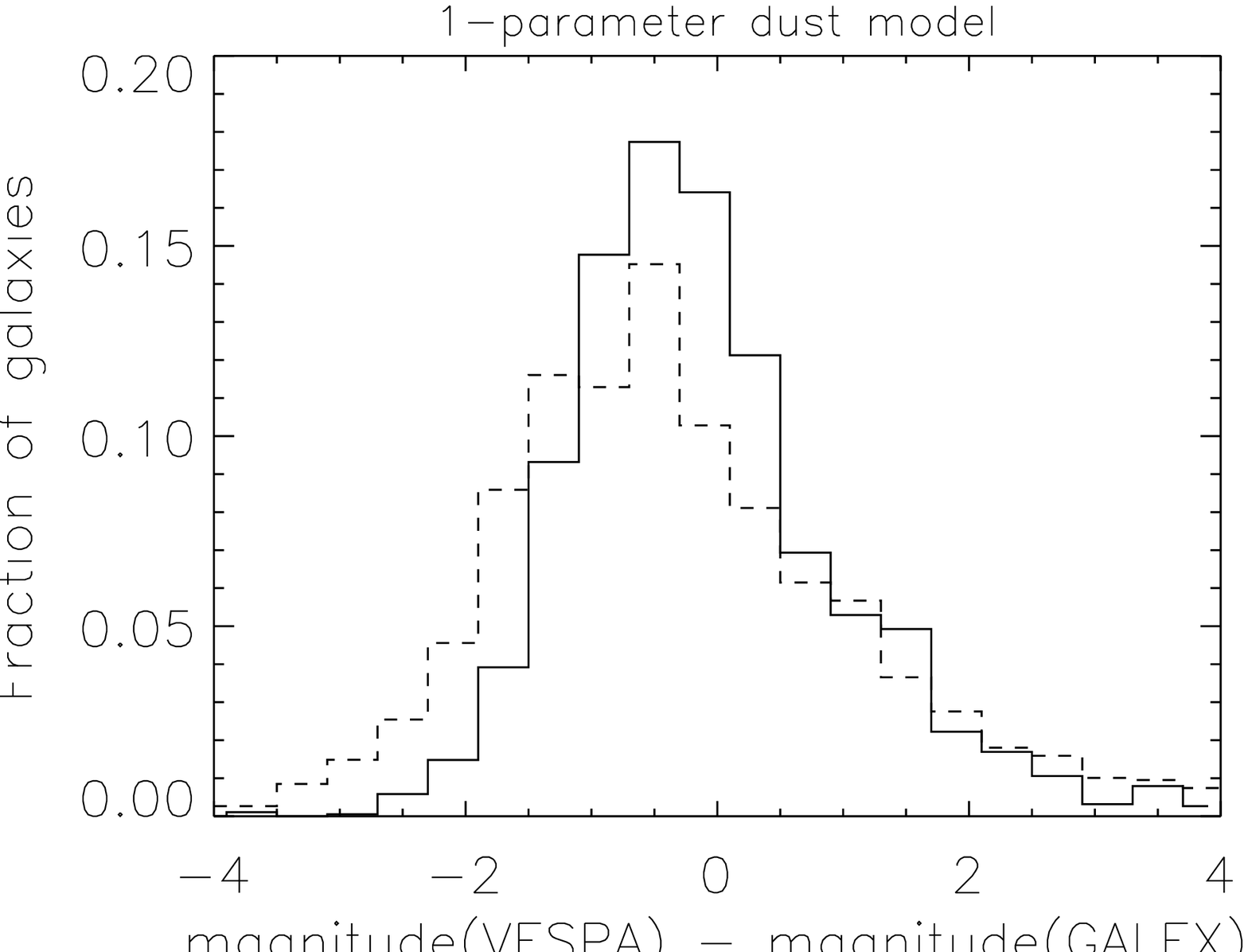}{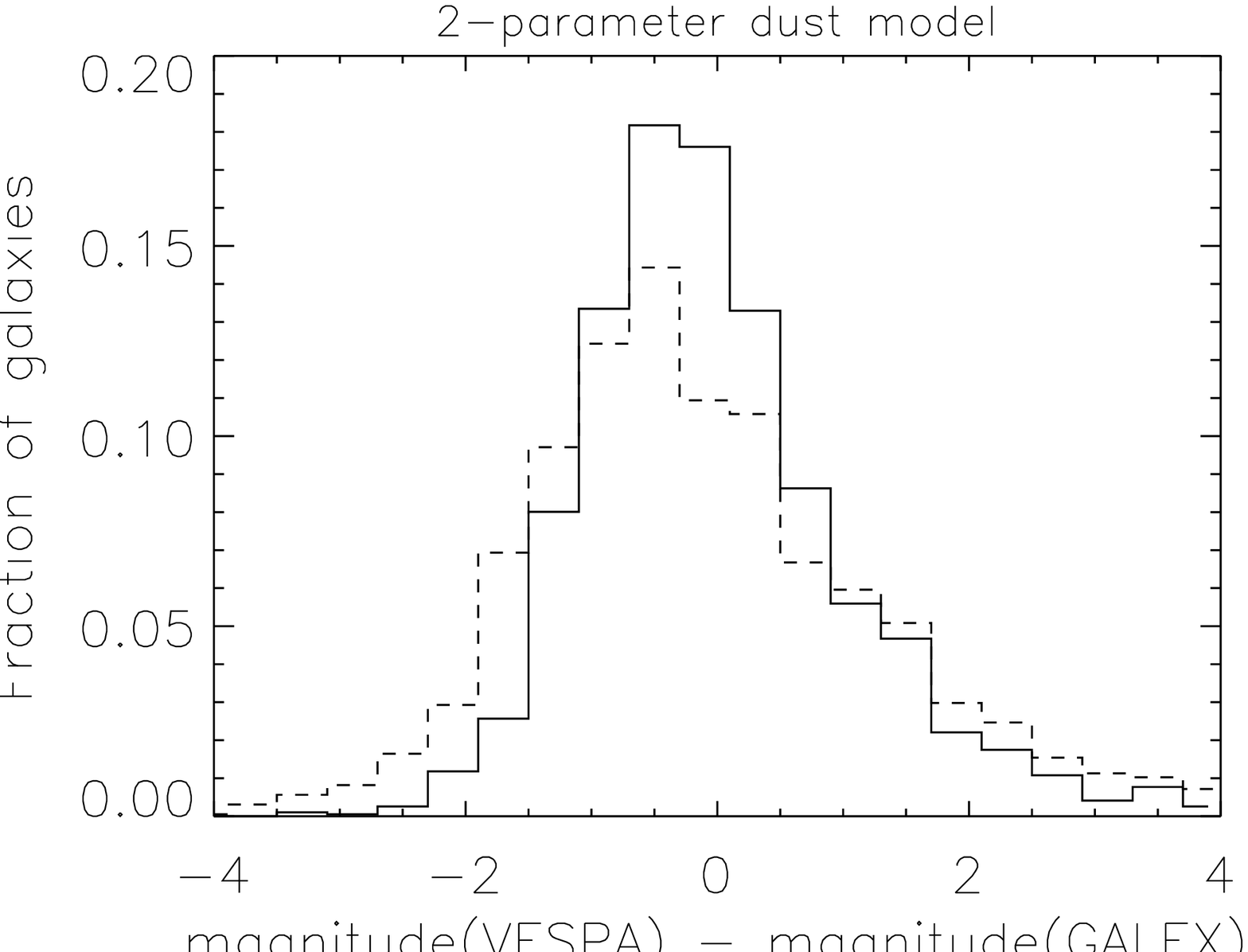}
\vspace{0.1in}
\caption{The distribution of the difference between the VESPA estimated GALEX magnitudes, and the observed magnitudes. In each panel the solid line refers to the near-UV filter, and the dashed line to the far-UV.}
\label{fig:galex_comparisons}
\end{figure*}

There are over 300,000 objects in the MGS which have GALEX magnitudes \citep{MartinEtAl05}. The SDSS optical range stops redwards of the GALEX two filters, and it is interesting to see how the predicted GALEX magnitudes from the recovered spectral fits compare to the observed ones. \\

As remarked earlier, Figure \ref{fig:UV_examples} shows how the recovered UV spectrum compares to the GALEX measured fluxes for nine random galaxies. Although not representative, it serves as a good visual indication that the UV flux might be a way to distinguish between the one- and the two-parameter dust model. For a more quantitative analysis,  we computed GALEX magnitudes from our recovered spectra and the filters transmission curves and compared them to the published GALEX magnitudes. Figure \ref{fig:galex_comparisons} shows the histogram of the difference of the two, for both GALEX filters and in the case of the two dust models. We note three things about these histograms: there is only a very small difference between the one- and two- parameter dust models; there is an offset of the peak of the two distributions with respect to zero; and the scatter in the far-UV filter is larger than in the near-UV filter. Let us now comment on each of these points in turn. \\

\begin{figure*}
\plottwo{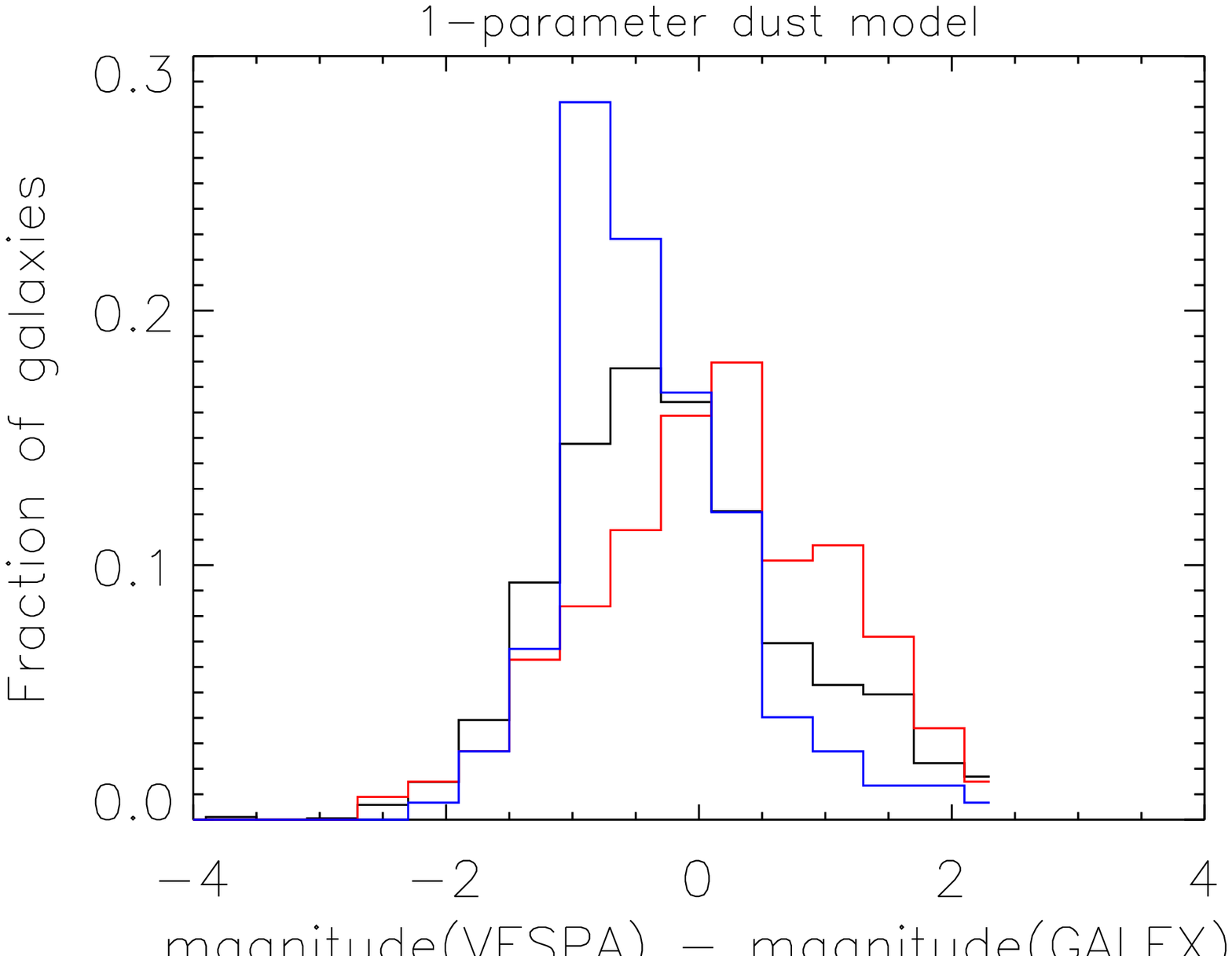}{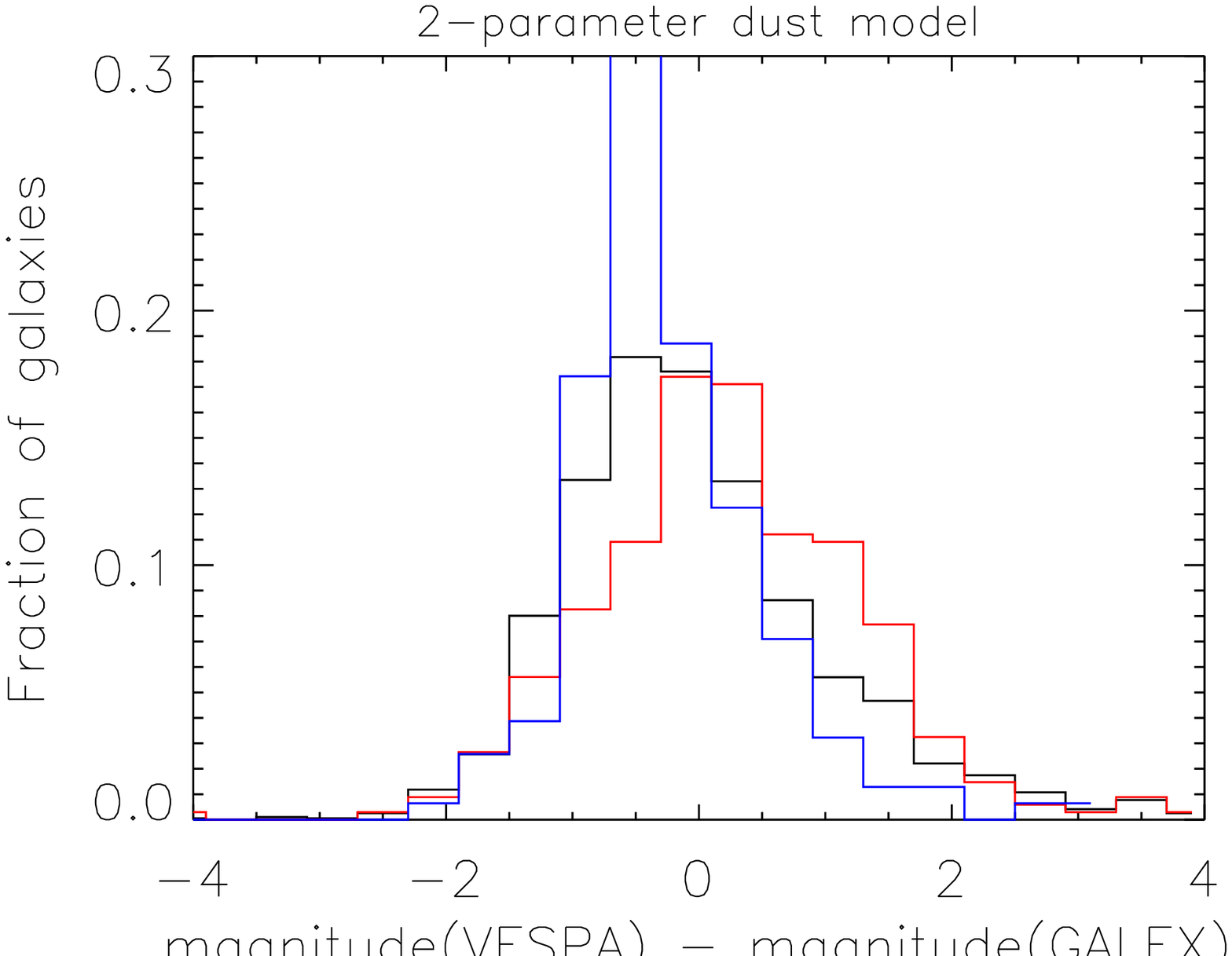}
\vspace{0.1in}
\caption{The distribution of the difference between the VESPA estimated GALEX magnitudes, and the observed magnitudes for the near-UV filter. In each panel the center black histogram is for all galaxies (the same as in Figure \ref{fig:galex_comparisons}), the red histogram (shifted to the right) is for red galaxies and the blue histogram (shifted to the left) is for blue galaxies.  }
\label{fig:galex_comparisons_red_blue}
\end{figure*}

The similarity of the two panels in Figure \ref{fig:galex_comparisons} suggests that perhaps on average both models do just as well. We take this one step further by repeating the analysis on red and blue galaxies separately, for the near-UV filter - the resulting histograms are shown in Figure \ref{fig:galex_comparisons_red_blue}. We find once again that the distributions from the two dust models are very similar, but there are differences associated with the color of the galaxy. Red galaxies have a larger scatter, but are not offset with respect to zero. We can also distinguish a slight peak in the histogram of red galaxies, where the VESPA magnitudes are fainter than the measured GALEX magnitudes by around one magnitude. This corresponds to cases where VESPA predicts no recent star formation, but GALEX tells us it is there. Visually, it corresponds to the case in the central panel of Figure \ref{fig:UV_examples}. This happens in a small number of galaxies, but tells us that there is valuable information to be added to the optical SDSS spectrum of a galaxy. The offset to brighter VESPA magnitudes we saw before comes mostly from blue galaxies, which leads us into the next point. \\

\begin{figure}
\plotone{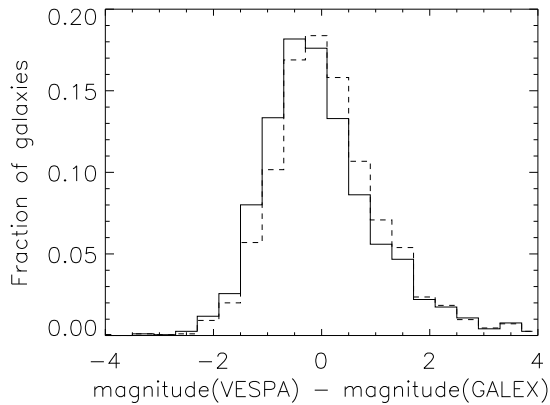}
\vspace{0.1in}
\caption{The distribution of the difference between the VESPA estimated GALEX magnitudes, and the observed magnitudes for the near-UV filter. The solid line assumes a $\lambda^{-0.7}$ dust law ((the same as in Figure \ref{fig:galex_comparisons}), and the dashed line assumes a steeper $\lambda^{-0.75}$ dust law but the same values of $\tau_V$ (see text for details).}
\label{fig:galex_nuv_alpha075_all}
\end{figure}

Our dust curve goes as $\lambda^{-0.7}$, which is a relatively flat extinction curve (see e.g. Figure \ref{fig:dust_curves}). A steeper curve would mean that VESPA magnitudes would be fainter (larger), and shift the histograms in Figure \ref{fig:galex_comparisons} closer to zero. To investigate how a steeper dust curve in the UV might affect these histograms, we corrected the VESPA magnitudes to a dust curve which goes as $\lambda^{-0.75}$ {\it assuming the same amount of dust}, i.e., keeping $\tau_V^{ISM}$ and $\tau_V^{BC}$ fixed. The resulting histogram, for the near-UV filter, can be seen in Figure \ref{fig:galex_nuv_alpha075_all}. The change comes almost exclusively from the blue galaxies, as expected. This suggests that our choice of dust law might be too flat in this wavelength range. This is relevant if we want to include GALEX data in the fits, as we intend to do in the future. \\

\begin{figure}
\plotone{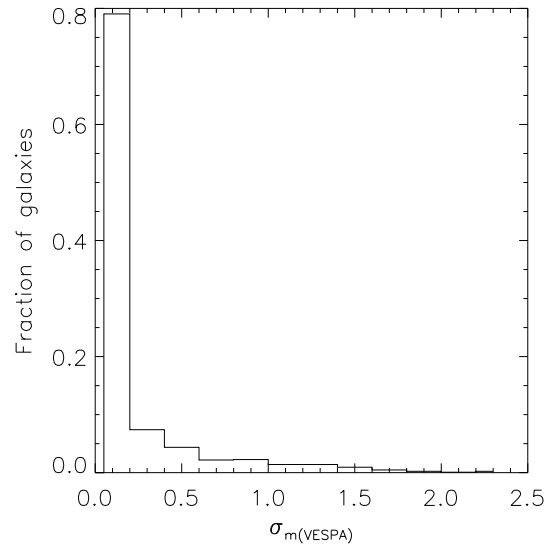}
\vspace{0.1in}
\caption{The error on recovered GALEX near-UV magnitudes, taking into account photon noise on the SDSS spectra only.}
\label{fig:galex_errors_vespa}
\end{figure}

Finally, we investigate whether the large scatter in the magnitude differences can be explained by the recovered errors in the mass fractions. This is particularly important when we are talking about the UV and young stars, as a small change in mass translates into a large difference in magnitudes. The error on each recovered flux point is

\begin{equation}
\sigma^2_{F_j} = \sum_{\alpha, \beta} S_{j,\alpha} C_{\alpha\beta}(m) S_{j,\beta} \end{equation}

where $C(m)$ is the covariance matrix for the estimated masses, corrected for recycling fractions and fiber aperture. From this the error on the estimated GALEX magnitude, $\sigma_m(VESPA)$ can be estimated. The distribution of these errors, for the near-UV, is shown in Figure \ref{fig:galex_errors_vespa}. \\ 

The histogram is similar for the far-UV filter, but the errors are generally larger. This makes perfect sense, as we expect the that region to be more sensitive to young stars and is a possible explanation for the slightly larger scatter for this filter in Figure \ref{fig:galex_comparisons}. However, it is obvious that $\sigma_m(VESPA)$ alone cannot explain the scatter in Figures \ref{fig:galex_comparisons} and \ref{fig:galex_comparisons_red_blue}. In practice this is not entirely surprising, given that the dust law alone will affect the estimated GALEX magnitudes from VESPA and this is a source of error not accounted for at the moment. Another important source of scatter is undoubtedly the fiber correction, which we calculate using the $z-$band. We are effectively extrapolating for the total flux from young stars using a correction based on old populations, and we expect some scatter between the two. \\


We conclude that {\it on average}, the star formation histories recovered by VESPA are accurate enough, even at young ages, to predict UV magnitudes that are far beyond the fitted spectral regime. However, the spectral range offered by the SDSS is simply not enough to break some degeneracies, especially concerning young star formation and dust. From a technical point of view, the extra information in GALEX photometry can be easily incorporated in the VESPA analysis, as can any other photometry at the other side of the spectral range.

\section{Conclusions} \label{sec:conclusion}

We presented a catalogue of star formation and metallicity histories, dust content and stellar masses for nearly 800,000 galaxies in SDSS' MGS and LRG samples, which we are now making public. The catalogue is the result of applying VESPA to SDSS' latest and final data release. VESPA has a self-regularization mechanism which gives an estimate of how many parameters one should recover from a given galaxy, given the quality of the data, and puts the emphasis on the robustness of the solutions, rather than highly-resolved star formations histories. \\

We find that the number of populations recovered from the MGS and the LRG sample is very different - whereas LRGs seem to only call for one to three populations given the current data quality, galaxies in the MGS justify a more complex model, with VESPA typically recovering between two and five populations.\\

We explored some basic properties of the catalogue, and showed that the derived quantities make physical sense. Looking at average star formation histories and using a simple color cut to separate red and blue galaxies, we saw that red galaxies are older and have less dust than their blue counterparts. By averaging the mass formed in the last 115 Gyrs in each galaxy, we computed a star formation rate which is in agreement with those calculated by \cite{BrinchmannEtAl04}. \\

The goal of this paper is not to determine which set of theoretical models best describe the Universe. However, we note that when looking at rest-frame averaged star-formation histories, the combination of M05 synthesis models with a one-parameter dust model seem to give the smoothest star formation history. This is an indication that the mass recovered in the first 8 bins of VESPA is particularly sensitive to the choice of dust modelling (for the first four) and the choice of SSP modelling (for the next four). The effects on global properties, such as total present-day stellar mass is small, as is the effect on recovered old stellar populations, but we recommend care when the user needs resolved histories in these time-scales. \\

We also compared how the recovered dust values compare to observed Balmer decrements, and found a re-assuring qualitative agreement which allows us to reliably separate galaxies with different degrees of dust extinction. By extrapolating the recovered spectrum into the UV, we estimated GALEX magnitudes for a selection of galaxies which have been observed with GALEX. In blue galaxies, with recent star formation, we found a systematic offset between the estimated and observed magnitudes, with VESPA predicting systematically brighter magnitudes. We found that a steeper dust law in the UV, proportional to $\lambda^{-0.75}$, would explain this offset. We showed that VESPA is good at predicting lack of star formation given the SDSS range, but there is a small population of galaxies for which GALEX sees star formation when we do not. \\

In a future publication we intend to harvest the extra information in GALEX and other surveys that extend the wavelength range of the SDSS spectra by explicitly adding photometry to the VESPA fits.\\

The catalogue is released with a variety of models. As we have stressed throughout the paper, our goal is not to immediately distinguish between different models but rather to reaffirm the need to take care when using the catalogue to make conclusions about the Universe: our answers are often, although not always, model dependent. Giving the user a tool to know when and how much to worry is a necessary step in the right direction. As new models become available to the community, we will increment the catalogue accordingly.\\


\section{Acknowledgments}
RT would like to thank Mike Read from the IfA Wide-Field Astronomy Unit and Gerard Lemson from the GAVO Millennium database for essential work and support in the development and implementation of the database. SW is supported by an STFC studentship.\\

    Funding for the SDSS and SDSS-II has been provided by the Alfred
    P. Sloan Foundation, the Participating Institutions, the National
    Science Foundation, the U.S. Department of Energy, the National
    Aeronautics and Space Administration, the Japanese Monbukagakusho,
    the Max Planck Society, and the Higher Education Funding Council
    for England. The SDSS Web Site is http://www.sdss.org/. The SDSS is managed by the Astrophysical Research Consortium for the Participating Institutions. The Participating Institutions are the American Museum of Natural History, Astrophysical Institute Potsdam, University of Basel, University of Cambridge, Case Western Reserve University, University of Chicago, Drexel University, Fermilab, the Institute for Advanced Study, the Japan Participation Group, Johns Hopkins University, the Joint Institute for Nuclear Astrophysics, the Kavli Institute for Particle Astrophysics and Cosmology, the Korean Scientist Group, the Chinese Academy of Sciences (LAMOST), Los Alamos National Laboratory, the Max-Planck-Institute for Astronomy (MPIA), the Max-Planck-Institute for Astrophysics (MPA), New Mexico State University, Ohio State University, University of Pittsburgh, University of Portsmouth, Princeton University, the United States Naval Observatory, and the University of Washington.

\bibliographystyle{astron}
\bibliography{my_bibliography}

\end{document}